\documentclass[confernce]{IEEEtran}
\IEEEoverridecommandlockouts

\usepackage{float}
\usepackage{caption}
\usepackage{algorithm}
\usepackage{graphicx}
\usepackage{tabularx} 
\usepackage{threeparttable}
\usepackage{amsmath}
\usepackage{placeins}
\usepackage[normalem]{ulem}
\usepackage{latexsym}
\usepackage{subcaption}
\usepackage{times}
\usepackage{enumerate}
\usepackage{xspace}
\usepackage{calc}
\usepackage{url}
\usepackage{booktabs}
\usepackage[english]{babel}
\usepackage{algorithm, algcompatible}
\usepackage[noend]{algpseudocode}
\usepackage{hhline}
\usepackage{multirow}
\usepackage{makecell} 
\usepackage{hyperref}
\hypersetup{
	citecolor=green
}
\usepackage{wrapfig}
\usepackage{csquotes}
\usepackage{multicol}
\usepackage{tikz}
\usepackage{xparse}

\NewDocumentCommand{\statcirc}{ O{#2} m }{%
	\begin{tikzpicture}
		\fill[#2] (0,0) circle (1.0ex); 
		\fill[#1] (0,0) -- (90:1ex) arc (90:270:1ex) -- cycle; 
	\end{tikzpicture}
}

\usepackage{etoolbox}
\usepackage{tkz-euclide}
\tikzset{%
	pics/sema/.style args={#1/#2/#3}{code={%
			\ifstrequal{#2}{0}{%
				\node[circle,minimum width=2mm,draw,fill=#1] {};
			}{%
				\tkzDefPoint(0,0){O}
				\tkzDrawSector[R,fill=#1](O,2mm)(90,90-#2)
				\tkzDrawSector[R,fill=#3](O,2mm)(90-#2,90-360)
			}
	}},
}

\newcommand{\sgn}{\ensuremath {\texttt{SGN}}{\xspace}}
\newcommand{\sgnkg}{\ensuremath {\texttt{SGN.Kg}}{\xspace}}
\newcommand{\sgnsig}{\ensuremath {\texttt{SGN.Sig}}{\xspace}}
\newcommand{\sgnver}{\ensuremath {\texttt{SGN.Ver}}{\xspace}}

\newcommand{\asgn}{\ensuremath {\texttt{ASGN}}{\xspace}}
\newcommand{\asgnkg}{\ensuremath {\texttt{ASGN.Kg}}{\xspace}}
\newcommand{\asgnsig}{\ensuremath {\texttt{ASGN.ASig}}{\xspace}}
\newcommand{\asgnagg}{\ensuremath {\texttt{ASGN.Agg}}{\xspace}}

\newcommand{\asgnaver}{\ensuremath {\texttt{ASGN.AVer}}{\xspace}}

\newcommand{\fsgn}{\ensuremath {\texttt{FSGN}}{\xspace}}
\newcommand{\fsgnkg}{\ensuremath {\texttt{FSGN.Kg}}{\xspace}}
\newcommand{\fsgnupd}{\ensuremath {\texttt{FSGN.Upd}}{\xspace}}
\newcommand{\fsgnsig}{\ensuremath {\texttt{FSGN.Sig}}{\xspace}}
\newcommand{\fsgnver}{\ensuremath {\texttt{FSGN.Ver}}{\xspace}}

\newcommand{\hysgn}{\ensuremath {\texttt{HYSGN}}{\xspace}}
\newcommand{\hysgnkg}{\ensuremath {\texttt{HYSGN.Kg}}{\xspace}}
\newcommand{\hysgnsig}{\ensuremath {\texttt{HYSGN.Sig}}{\xspace}}

\newcommand{\hysgnver}{\ensuremath {\texttt{HYSGN.Ver}}{\xspace}}

\newcommand{\kg}{\ensuremath {\texttt{Kg}}{\xspace}}
\newcommand{\agg}{\ensuremath {\texttt{Agg}}{\xspace}}
\newcommand{\upd}{\ensuremath {\texttt{Upd}}{\xspace}}
\newcommand{\ssig}{\ensuremath {\texttt{Sig}}{\xspace}}
\newcommand{\asig}{\ensuremath {\texttt{ASig}}{\xspace}}
\newcommand{\ver}{\ensuremath {\texttt{Ver}}{\xspace}}
\newcommand{\aver}{\ensuremath {\texttt{AVer}}{\xspace}}
\newcommand{\comconstr}{\ensuremath {\texttt{ComConstr}}{\xspace}}

\newcommand{\Breakin}{\ensuremath {\texttt{Break-In(j)}}{\xspace}}
\newcommand{\comconstrmsk}{\ensuremath {\texttt{ComConstr}_{msk}(.)}{\xspace}}
\newcommand{\Sig}{\ensuremath{\texttt{Sig}_{sk}(.)}}
\newcommand{\Sigj}{\ensuremath{\texttt{Sig}_{sk_j}(.)}}

\newcommand{\hyhamu}{\ensuremath {\texttt{HYHAMU-SGN}}{\xspace}}

\newcommand{\hamusgn}{\ensuremath {\texttt{HAMU-SGN}}{\xspace}}
\newcommand{\hamusgnkg}{\ensuremath {\texttt{HAMU-SGN.Kg}}{\xspace}}

\newcommand{\hamusgnsig}{\ensuremath {\texttt{HAMU-SGN.Sig}}{\xspace}}

\newcommand{\hamusgnver}{\ensuremath {\texttt{HAMU-SGN.Ver}}{\xspace}}

\newcommand{\hamusgncomconst}{\ensuremath {\texttt{HAMU-SGN.ComConstr}}{\xspace}}

\newcommand{\hamuasgn}{\ensuremath {\texttt{HAMU-ASGN}}{\xspace}}
\newcommand{\hamuasgnkg}{\ensuremath {\texttt{HAMU-ASGN.Kg}}{\xspace}}

\newcommand{\hamuasgnsig}{\ensuremath {\texttt{HAMU-ASGN.Sig}}{\xspace}}
\newcommand{\hamuasgnasig}{\ensuremath {\texttt{HAMU-ASGN.ASig}}{\xspace}}
\newcommand{\hamuasgnver}{\ensuremath {\texttt{HAMU-ASGN.Ver}}{\xspace}}

\newcommand{\hamufsgn}{\ensuremath {\texttt{HAMU-FSGN}}{\xspace}}
\newcommand{\hamufsgnkg}{\ensuremath {\texttt{HAMU-FSGN.Kg}}{\xspace}}

\newcommand{\hamufsgnver}{\ensuremath {\texttt{HAMU-FSGN.Ver}}{\xspace}}

\newcommand{\sch}{\ensuremath {\texttt{Schnorr}{\xspace}}}
\newcommand{\schkg}{\ensuremath {\texttt{Schnorr.Kg}}{\xspace}}
\newcommand{\schsig}{\ensuremath {\texttt{Schnorr.Sig}}{\xspace}}
\newcommand{\schver}{\ensuremath {\texttt{Schnorr.Ver}}{\xspace}}

\newcommand{\hors}{\ensuremath {\texttt{HORS}{\xspace}}}
\newcommand{\horskg}{\ensuremath {\texttt{HORS.Kg}}{\xspace}}
\newcommand{\horssig}{\ensuremath {\texttt{HORS.Sig}}{\xspace}}
\newcommand{\horsver}{\ensuremath {\texttt{HORS.Ver}}{\xspace}}

\newcommand{\hases}{\ensuremath {\texttt{HASES}{\xspace}}}

\newcommand{\hasessig}{\ensuremath {\texttt{HASES.Sig}}{\xspace}}

\newcommand{\pqhases}{\ensuremath {\texttt{PQ-HASES}{\xspace}}}
\newcommand{\pqhaseskg}{\ensuremath {\texttt{PQ-HASES.Kg}}{\xspace}}
\newcommand{\pqhasessig}{\ensuremath {\texttt{PQ-HASES.Sig}}{\xspace}}
\newcommand{\pqhasesver}{\ensuremath {\texttt{PQ-HASES.Ver}}{\xspace}}

\newcommand{\pqhasescomconstr}{\ensuremath {\texttt{PQ-HASES.ComConstr}}{\xspace}}

\newcommand{\lahases}{\ensuremath {\texttt{LA-HASES}{\xspace}}}
\newcommand{\lahaseskg}{\ensuremath {\texttt{LA-HASES.Kg}}{\xspace}}
\newcommand{\lahasessig}{\ensuremath {\texttt{LA-HASES.ASig}}{\xspace}}
\newcommand{\lahasesver}{\ensuremath {\texttt{LA-HASES.AVer}}{\xspace}}
\newcommand{\lahasesagg}{\ensuremath {\texttt{LA-HASES.Agg}}{\xspace}}
\newcommand{\lahasescomconstr}{\ensuremath {\texttt{LA-HASES.ComConstr}}{\xspace}}

\newcommand{\hyhases}{\ensuremath {\texttt{HY-HASES}{\xspace}}}
\newcommand{\hyhaseskg}{\ensuremath {\texttt{HY-HASES.Kg}}{\xspace}}
\newcommand{\hyhasessig}{\ensuremath {\texttt{HY-HASES.Sig}}{\xspace}}

\newcommand{\hyhasesver}{\ensuremath {\texttt{HY-HASES.Ver}}{\xspace}}
\newcommand{\hyhasescomconstr}{\ensuremath {\texttt{HY-HASES.ComConstr}}{\xspace}}

\newcommand{\cco}{\ensuremath {\texttt{CCO}}{\xspace}}
\newcommand{\xmssmt}{\ensuremath {\text{XMSS}^\text{MT}{\xspace}}}

\newcommand{\Areal}{\ensuremath {\overrightarrow{A}_{\mathit{real}}}{\xspace}}
\newcommand{\Asim}{\ensuremath {\overrightarrow{A}_{\mathit{sim}}}{\xspace}}
\newcommand{\Ra}{\ensuremath \stackrel{\$}{\leftarrow}{\xspace}}
\newcommand{\Rq}{\ensuremath \stackrel{\$}{\leftarrow}\mathbb{Z}_{q}^{*}{\xspace}}

\newcommand{\as}{\ensuremath {\leftarrow}{\xspace}}
\newcommand{\prf}{\ensuremath {\texttt{PRF}}{\xspace}}

\newcommand{\msk}{\ensuremath {\mathit{msk}}{\xspace}}
\newcommand{\sk}{\ensuremath { \mathit{sk} }{\xspace}}
\newcommand{\pk}{\ensuremath { \mathit{PK} }{\xspace}}
\newcommand{\csk}{\ensuremath { \mathit{csk} }{\xspace}}
\newcommand{\cpk}{\ensuremath { \mathit{CPK} }{\xspace}}

\newcommand{\dlp}{\ensuremath { \texttt{DLP} }{\xspace}}
\newcommand{\dl}{\ensuremath {\mathit{DL}}{\xspace}}
\newcommand{\advAdl}{\ensuremath {\mathit{Adv}_{G}^{\dl}(\mathcal{A})}{\xspace}}
\newcommand{\advdl}{\ensuremath {\mathit{Adv}_{G}^{\dl}(t)}{\xspace}}
\newcommand{\advdll}{\ensuremath {\mathit{Adv}_{G, \alpha}^{\dl}(t')}{\xspace}}

\newcommand{\EUCMA}{\ensuremath { \texttt{EU-CMA} }{\xspace}}

\newcommand{\AEUCMA}{\ensuremath { \texttt{A-EU-CMA} }{\xspace}}
\newcommand{\FEUCMA}{\ensuremath { \texttt{F-EU-CMA} }{\xspace}}
\newcommand{\HEUCMA}{\ensuremath { \texttt{H-EU-CMA} }{\xspace}}

\newcommand{\ro}{\ensuremath {\mathit{RO}(.)}{\xspace}}

\newcommand{\hsim}{\ensuremath {\mathit{H}\mhyphen\mathit{Sim}}{\xspace}}

\newcommand{\lh}{\ensuremath {\mathcal{LH}}{\xspace}}
\newcommand{\lm}{\ensuremath {\mathcal{LM}}{\xspace}}
\newcommand{\ls}{\ensuremath {\mathcal{LS}}{\xspace}}
\newcommand{\lr}{\ensuremath {\mathcal{LR}}{\xspace}}
\newcommand{\lc}{\ensuremath {\mathcal{LC}}{\xspace}}
\newcommand{\RNG}{\ensuremath { \texttt{RNG} }{\xspace}}

\newcommand{\nab}{\ensuremath {\overline{\mathit{E1}}}{\xspace}}
\newcommand{\forge}{\ensuremath {\mathit{E2}}}{\xspace}
\newcommand{\nabb}{\ensuremath {\mathit{\overline{E3}}}{\xspace}}
\newcommand{\suc}{\ensuremath {\mathit{Win}}{\xspace}}

\newcommand{\concat}{\mathbin\Vert}

\newcommand{\advhysgn}{\ensuremath {\mathit{Adv}_{\hysgn}^{\HEUCMA}(t,q_s,q_s')}{\xspace}}

\newcommand{\advhamuagg}{\ensuremath {\mathit{Adv}_{\mathit{\hamuasgn}}^{\AEUCMA}(t, q_s, q_s')}{\xspace}}
\newcommand{\advhamufs}{\ensuremath {\mathit{Adv}_{\mathit{\hamufsgn}}^{\FEUCMA}(t, q_s, q_s',1)}{\xspace}}
\newcommand{\advhyhases}{\ensuremath {\mathit{Adv}_{\mathit{\hyhases}}^{\HEUCMA}(t,q_s, q_s')}{\xspace}}
\newcommand{\advpqhases}{\ensuremath {\mathit{Adv}_{\mathit{\pqhases}}^{\FEUCMA}(t,q_s, q_s', 1)}{\xspace}}
\newcommand{\advlahases}{\ensuremath {\mathit{Adv}_{\mathit{\lahases}}^{\AEUCMA}(t,q_s, q_s')}{\xspace}}
\newcommand{\advhors}{\ensuremath {\mathit{Adv}_{\mathit{HORS}}^{\EUCMA}(t',q_s, q_s')}{\xspace}}
\newcommand{\advhyhors}{\ensuremath {\mathit{Adv}_{\mathit{HORS}}^{\EUCMA}(t_{PQ}',q_s. q_s')}{\xspace}}
\newcommand{\advhydll}{\ensuremath {\mathit{Adv}_{G, \alpha}^{\dl}(t_{LA}')}{\xspace}}

\newcommand{\A}{$\mathcal{A}$}
\newcommand{\F}{$\mathcal{F}$}

\newcommand\tab[1][1cm]{\hspace*{#1}}
\newcommand{\algrule}[1][.2pt]{\par\vskip.5\baselineskip\hrule height #1\par\vskip.5\baselineskip}
\newcommand{\specialcell}[2][c]{
	\begin{tabular}[#1]{@{}c@{}}#2\end{tabular}}

\newcounter{algsubstate}
\renewcommand{\thealgsubstate}{\alph{algsubstate}}

\usepackage{mathtools, stmaryrd}
\usepackage{xparse} \DeclarePairedDelimiterX{\Iintv}[1]{\llbracket}{\rrbracket}{\iintvargs{#1}}
\NewDocumentCommand{\iintvargs}{>{\SplitArgument{1}{,}}m}
{\iintvargsaux#1} %
\NewDocumentCommand{\iintvargsaux}{mm} {#1\mkern1.5mu,\mkern1.5mu#2}

\newcommand{\romannum}[1]{\uppercase\expandafter{\romannumeral #1\relax}}

\mathchardef\mhyphen="2D 

\newcolumntype{P}[1]{>{\centering\arraybackslash}p{#1}}
\newcolumntype{M}[1]{>{\centering\arraybackslash}m{#1}}

\usepackage{theorem}

\urldef\myurl\url{http://www4.ncsu.edu/}
\urldef\miracle\url{http://www.shamus.ie/}

\newtheorem{theorem}{Theorem}
{\theorembodyfont{\rmfamily}
	{\bfseries}{\rmfamily}}
{\theorembodyfont{\rmfamily}
	\newtheorem{definition}{Definition}{\bfseries}{\rmfamily}}
{\theorembodyfont{\rmfamily}
	{\bfseries}{\rmfamily}}
{\bfseries}{\rmfamily}

\newtheorem{theoremalt}{Theorem}[theorem]

\newenvironment{theoremp}[1]{
  \renewcommand\thetheoremalt{#1}
  \theoremalt
}{\endtheoremalt}

\usepackage{cite}
\usepackage{amsmath,amssymb,amsfonts}
\usepackage{graphicx}
\usepackage{textcomp}
\usepackage{xcolor}
\def\BibTeX{{\rm B\kern-.05em{\sc i\kern-.025em b}\kern-.08em
    T\kern-.1667em\lower.7ex\hbox{E}\kern-.125emX}}

\usepackage[
top    = 0.75in,
bottom = 1in,
left   = 0.625in,
right  = 0.625in]{geometry}

	\title{Post-Quantum Hybrid Digital Signatures with Hardware-Support for Digital Twins}

\author{ Saif E.~Nouma  and	Attila A.~Yavuz  \\
	Department of Computer Science and Engineering \\
	University of South Florida \\
	Tampa, FL \\
	\{saifeddineouma, attilaayavuz\}@usf.edu \\
}

\date{}
\usepackage{url}

\begin{document}
	\maketitle
	
	\vspace{-12pt}
	
\begin{abstract}

Digital Twins (DT) virtually model cyber-physical objects using Internet of Things (IoT) components (e.g., sensors) to gather and process senstive information stored in the cloud. Trustworthiness of the streamed data is crucial which requires quantum safety and breach resiliency. Digital signatures are essential for scalable authentication and non-repudiation. Yet, NIST PQC signature standards are exorbitantly costly for low-end IoT without considering forward security. Moreover, Post-Quantum (PQ) signatures lack aggregation, which is highly desirable to reduce the transmission and storage burdens in DTs. Hence, there is an urgent need for lightweight digital signatures that offer compromise resiliency and compactness while permitting an effective transition into the PQ era for DTs.

We create a series of highly lightweight digital signatures called {\em Hardware-ASsisted Efficient Signature} (\hases) that meets the above requirements. The core of \hases~is a hardware-assisted cryptographic commitment construct oracle (\cco) that permits verifiers to obtain expensive commitments without signer interaction. We created three \hases~schemes: \pqhases~is a forward-secure PQ signature, \lahases~is an efficient aggregate Elliptic-Curve signature, and \hyhases~is a novel hybrid scheme that combines \pqhases~and \lahases~with novel strong nesting and sequential aggregation. \hases~does not require a secure-hardware on the signer. We proved that \hases~schemes are secure and implemented them on commodity hardware and an 8-bit AVR ATmega2560. Our experiments confirm that \pqhases~and \lahases~are two magnitudes of times more signer efficient than their PQ and conventional-secure counterparts, respectively. \hyhases~outperforms NIST PQC and conventional signature combinations, offering a standard-compliant transitional solution for emerging DTs. We open-source \hases~schemes for public testing and adaptation.

\end{abstract}


	\section{Introduction} \label{sec:Introduction}


Digital Twins (DT)  paradigm aims to represent a digital replica of the physical systems \cite{aloqaily2022integrating}. It can facilitate the means to monitor, understand, and optimize the functions of physical entities living, or non-living \cite{el2021potential}. They are primarily empowered by Internet of Things (IoT) components (e.g., smart sensors, actuators) to approximate the behavior of twins.  Hence, DTs can play an important role in many real-life applications such as constructing parallel metaverse world. In particular, DTs are suitable for IoT-cloud enabled systems with human-centric applications \cite{shengli2021human}.   For instance, IoT devices of a patient (e.g., medical implants, haptic sensors)  constantly gather highly security sensitive information to be securely offloaded to a remote cloud/edge server for data analytics and smart decision-making \cite{aloqaily2022integrating}. Subsequently, such medical DT can aid physicians in monitoring the patient's well-being and contribute to the prevention of potential illnesses from manifesting or advancing.   

DT applications gather, process and offload a large volume of heterogeneous data streams, which entail potentially security-sensitive information (e.g., healthcare, financial, personal). Therefore, it is imperative to ensure the trustworthiness of sensitive data maintained by the DT-Cloud continuum.  Especially, authentication and integrity are foundational security measures for safeguarding DT systems against a range of potential attacks (e.g., data tampering, fraud, man-in-middle). Digital signatures~\cite{Ed25519} provide scalable authentication/integrity as well as non-repudiation and public verifiability via public-key infrastructures. Hence, they are essential primitives to ensure security and trust for DTs. Yet, DT applications have security and performance requirements that are well beyond what conventional (standard) signatures can offer. Below, we outline some of the important properties that a digital signature should offer to be a viable solution for emerging DTs.

\vspace*{1mm}
{\em (i) Post-Quantum (PQ) Security}: With the anticipated arrival of quantum computers~\cite{chamola2021information}, Shor's algorithm \cite{Shor-algo} can break conventional cryptosystems (e.g., ECDSA~\cite{Ed25519}, RSA~\cite{katz2020introduction}) that rely on conventional intractability assumptions (e.g., Discrete Logarithm Problem (DLP)~\cite{katz2020introduction}). The fast progress in quantum computing make it necessary to swiftly transition to quantum-secure alternatives. Indeed, NIST initiated PQC standardization, outlining selected PQ signature standards. The White House also issued a memorandum providing guidance for the transition to PQC \footnote{\url{https://www.whitehouse.gov/wp-content/uploads/2022/11/M-23-02-M-Memo-on-Migrating-to-Post-Quantum-Cryptography.pdf}}. 
Thus, DT applications urgently need a PQ digital signature that respects their efficiency requirements.

{\em (ii) Compromise-resiliency}: Most of the DT applications such as smart-health monitoring or smart-home, sensors, actuators, and mobile devices may operate in open and hostile environments that make them vulnerable to compromise via either physical means or malware \cite{wang2021quantum2fa}. Hence, it is important for a digital signature to offer compromise-resiliency features like forward security (FS) \cite{shaw2022post}. 
The latter prevents forgeries for the past time periods upon private key exposure events.
Hence, it guarantees authenticity and integrity before a system breach point \cite{ForwardSecure_MMM_02} by constantly updating the secret keys. FS signatures are (in some cases significantly) costlier than their standard (plain) signature counterparts. This becomes even more expensive when PQ security is also considered.

{\em (iii) Signer Efficiency:} DT systems harbor large quantities of low-end IoT devices (e.g., smart sensors) that are highly resource-limited (e.g., battery, CPU, memory, bandwidth).  Hence, it is crucial to minimize the overhead of digital signatures over DT systems to permit a secure yet practical deployment. Hence, the digital signature should offer computationally lightweight operations and small signature sizes to reduce the impact of cryptographic overhead on battery life. Yet, even conventional signature standards (e.g., ECDSA) are considered expensive for some IoT applications. More importantly, the NIST PQC signatures are infeasible on low-end platforms, and currently impractical with advanced features like forward security.


%
%

{\em (iv) Ease of Transitioning and Cryptographic Agility}: 
The transition to PQC is a challenging and error-prone progress due to the need to update current cryptographic implementations. Several factors need to be considered, including cryptographic solutions for heterogeneous hardware devices (e.g., IoT devices). Thus, it is highly desirable to adopt PQC that offers backward compatibility with prior implementations, which can facilitate the implementation and transition process. 
For instance, hash-based schemes utilize cryptographic hash functions (e.g., SHA-2), which are already present in all standard-compliant cryptographic software libraries. Moreover, hash-based algorithms are considered PQ-safe with minimal intractability assumptions.

NIST emphasizes the need for a {\em hybrid PQ signature} that combines both classical and post-quantum variants~\cite{barker2018recommendation}. The hybrid signature offers the advantage that  as long as at least one of the algorithms used remains secure, the hybrid scheme will remain secure~\cite{bindel2017transitioning}. It not only enhances security definitions by requiring adversaries to break each algorithm but also allows for the utilization of existing libraries of traditional cryptosystems while fulfilling some PQC. This concept promotes {\em cryptographic agility}~\cite{joseph2022transitioning}, which involves designing protocols that can support multiple algorithms simultaneously, as a safety and security measure against the vulnerability of deployed ones. 

\subsection{Related Work and Limitations of the State-of-the-Art} \label{subsec:RelatedWork}

We focus on digital signatures with desirable properties for DT use-case. First, we discuss special signature schemes and their constraints, followed by complementary related works.

\noindent \textbf{Digital Signatures with Special Properties.} We discuss relevant signatures to ours with an emphasis on PQC, compromise-resiliency, compactness (aggregate) and finally hybrid features.  
\vspace{1pt}

$\bullet$ {\em Post-Quantum (PQ) Digital Signatures:} As part of the NIST PQC standardization efforts~\cite{nist-4th-round}, the selected PQ signature standards include two lattice-based schemes, Dilithium \cite{ducas2018crystals} and Falcon \cite{fouque2018falcon}, and a hash-based SPHINCS+ \cite{bernstein2019sphincs+}.  SPHINCS+ incurs high energy and bandwidth overheads due to expensive signature generation and large key sizes, making it infeasible for frequent data authentication on low-end devices. The two lattice-based signatures offer a better balance between the performance metrics (i.e., storage, transmission, computation), but are based on new intractability assumptions, compared to hash-based schemes. 
To our knowledge, the implementation of NIST standards still does not support the resource-constrained devices (e.g., 8-bit microcontrollers (MCUs)) which are widely deployed in the IoTs. For example, implementing Falcon \cite{yavuz2022distributed} on such devices is challenging as it requires double-precision floating-point operations, which are not supported by many low-end devices. The lattice-based BLISS is the only PQ signature with benchmarking on an 8-bit device but vulnerable to numerous side-channel and timing attacks \cite{espitau2017side}.

$\bullet$ {\em Forward-Secure (FS) Digital Signatures:} offer improved protection against system breaches (e.g., malware attacks) by employing a key update strategy \cite{chen2022lfs}. The frequent private key update results in increased size of public keys. A straightforward technique consists of refreshing the private key via a one-way hash function which results in a linear public key blow-up w.r.t. the number of messages to be signed \cite{Yavuz:2012:TISSEC:FIBAF}. This results in a linear bandwidth overhead at the signer which is deemed to be a resource-constrained device. Thus, it is not scalable to the large DT environments. On the other side, there exist generic transformations (e.g., \cite{yavuz2022frog, Yavuz:CORE:CNS:2020}) that introduce forward security property to a normal digital signature. The most efficient approach introduces a $log_2(J)$ signing overhead blow-up, where $J$ is the maximum number of messages to be signed. The extra overhead exacerbates when PQ security is considered. For example, the RFC standard and NIST recommendation, \xmssmt~\cite{cooper2020recommendation} is currently the only PQ and FS signature standard. Its signature generation is more than magnitudes times costlier than the NIST PQC standard Dilithium. Another line of research (e.g., a lattice-based FS-PQ digital signature ANT \cite{behnia2021towards}) enables signers to delegate the public-key computation to a set of distributed servers. Despite their efficient signing and compact key sizes, such approaches assume non-colluding servers that not only risky for some DT applications but also introduces heavy network delays. Additionally, there exist a recent identity-based digital signature with forward security \cite{shaw2022post}. However, it does not offer a lightweight signing and no benchmark on low-end device is reported.

$\bullet$ {\em Aggregate Digital Signatures:}  Aggregate Signature (AS) are vital for frequent and continuous authentication, particularly on low-end devices and bandwidth-limited networks. Formally, an AS scheme reduces the cryptographic overhead by combining multiple distinct signatures, that can be issued from a single or multiple signers, into a fixed-length signature. Thus, aggregation improve the performance efficiency by lowering the cryptographic payload into a constant overhead. ASs have seen numerous applications, namely in IoT networks \cite{bagchi2023post}, delay-tolerant networks \cite{zhu2009smart}, secure routing \cite{boldyreva2007ordered}, and distributed systems (e.g., Blockchain) \cite{drijvers2020pixel}.

Several aggregate signatures have been proposed in the literature which can be divided into:
{\em (i) Factorization-based:} rely on the hardness assumption of the factorization problem. For example, the condensed RSA (C-RSA) \cite{Yavuz:TDSC:OutsourcedDB} signature offers aggregation with an efficient verification but expensive signing and large key sizes due to the modular exponentiation over large prime modulus. 
{\em (ii) Pairing-based:} based on bilinear maps and offer aggregation within multi-user settings. BLS \cite{BLS:2004:Boneh:JournalofCrypto} represents the widely-used seminal pairing-based signature which offers a small signature and public key sizes, but with a costly pairing and map-to-point operations that are highly expensive.
Recent optimized pairing-based aggregate signatures are back-traced to BLS in terms of signer efficiency. For instance, LFS-AS \cite{chen2022lfs} is a pairing-based aggregate signature with FS for e-Health. Its signature generation performs four exponentiation which are expensive for 8-bit devices. Its verification is even more costlier since due to pairing operations. Numerous aggregate signatures with advanced properties (e.g., certificateless setting) \cite{vallent2021efficient} are pairing-free schemes. However, such recent works focus on reducing the verification computational cost while omitting the prohibitive signing overhead for 8-bit MCUs.
{\em (iii)} Elliptic Curve (EC)-based: The aggregate EC-based schemes offer a balance in terms of signature generation efficiency and key sizes. 
For instance, FI-BAF \cite{Yavuz:2012:TISSEC:FIBAF} is a signer-efficient EC-based AS and FS digital signature. However, it incurs persistent linear storage overhead at verifiers, wherein one-time public keys must be periodically redistributed. Thus, it is not scalable for large-scale data-intensive DT applications. There exist recent EC-based AS schemes with advanced properties for IoTs (e.g., \cite{vallent2021efficient, li2020permissioned}). However, they inherit similar signing efficiency to that of Schnorr \cite{SchnorrQ} at the signer side and they lack low-level implementation on resource-constrained devices. 

There is very limited work on PQ aggregate signatures. For example, a recent lattice-based AS scheme \cite{boneh2020one} only works for non-interactive one-time or interactive multiple-time settings, with a logarithmic compression rate w.r.t the number of signatures. Another approach converts classical DLP-based aggregate signatures into lattice dimensions using the Fiat-Shamir with Aborts paradigm \cite{boudgoust2023sequential}, but this method requires additional computational overhead and again very low compression ratios. Additionally, there exist numerous lattice-based aggregate signatures, optimized for blockchain applications (e.g., \cite{bagchi2023post}). Again, they have costly signature generation with the absence of benchmark on low-end IoT.  Therefore, PQ-secure ASs are currently not feasible for our envisioned applications.

$\bullet$ {\em Hybrid Approaches:} 
Various hybrid digital signatures aim to establish robust authentication methods by combining different digital signatures and other emerging technologies (e.g., quantum networks). For instance, \cite{yavuz2022distributed} proposed a hybrid protocol that incorporates NIST PQC standards, quantum networks, and hardware acceleration methods, thus enabling swift and quantum-safe execution of distributed protocols.
There exist hybrid key agreement protocols (e.g., \cite{qassim2017post}) that employ post-quantum cryptography and physical layer methods, which are orthogonal to ours. 
Crocket et al. \cite{crockett2019prototyping} integrate PQ and hybrid digital signatures into Internet security protocols (e.g., TLS). Those works are complementary to ours. 
Paul et al. \cite{paul2020towards} benchmark a range of hybrid digital signatures by combining NIST standards including conventional (e.g., RSA) and PQ (e.g., Falcon) signature schemes. However, our findings reveal that both categories involve costly signing operations, making them unsuitable for low-end IoT even before considering advanced security properties. Notably, there is a gap in hybrid signature solutions that effectively address quantum threats while also embodying the desirable attributes of conventional signatures. NIST underscores the significance of hybrid conventional-PQ-secure cryptosystems \cite{barker2018recommendation} to offer fail-safe designs against unexpected failure of emerging PQC schemes \cite{ott2019identifying,bindel2017transitioning}, while also retaining cryptgoraphic agility \cite{joseph2022transitioning}.

\vspace{2mm}
\noindent \textbf{Other/Complementary Related Work.}

$\bullet$ {\em Secure Hardware-Assisted Primitives}: Another line of research exploits the availability of trusted execution environments in modern architectures to achieve higher cryptographic functionalities. For instance, it is possible to emulate asymmetric cryptosystems from symmetric-key algorithms (e.g., MACs) with a secure enclave (e.g., Intel Software Guard (SGX) ~\cite{el2022benchmarking}). While efficient and foreseen to be PQ-secure, it requires each party to have a local secure enclave (e.g., SCB~\cite{ouyang2021scb}), which is not practical for low-end devices that represent a major part of DTs. Moreover, they provide restricted public verifiability (only SGX-enabled devices) and lack non-repudiation (due to shared symmetric keys), which is a critical need for numerous real-life applications.

$\bullet$ {\em Other Special Authentication Techniques:}  
Proof of Data Possession (PDP) \cite{ateniese2008scalable} and Proof of Retrievability (PoR)~\cite{anthoine2021dynamic} protocols can offer public auditing of the outsourced user data. They offer fast verification time of the authenticated data via interactive random checks. They mostly rely on homomorphic linear authenticators (HLA) \cite{wang2011privacy} which allows auditors to perform verification without retrieving the entire data. Most HLAs are also implemented via the pairing-based or RSA-based schemes.
Despite their merits, most of  the above approaches rely on foundational operations/signatures such as  EC scalar multiplications (e.g., Ed25519), RSA, or BLS signatures at the signer. Moreover, we observe the absence of performance evaluations on low-end devices (e.g., 8-bit ATMega2560). In our comparisons, we focus on Ed25519~\cite{Ed25519}, RSA \cite{katz2020introduction}, and BLS \cite{BLS:2004:Boneh:JournalofCrypto} to represent the signer overhead of the schemes that rely on such operations (see Section \ref{sec:performance_analysis}).  



\subsection{Our Contribution}

We created a new series of highly lightweight digital signatures called {\em HArdware-Assisted Efficient  Signatures (\hases)}. \hases~efficiently combines above seemingly conflicting attributes while ensuring near-optimal signer performance to meet resource-constrained DT requirements. Our main component is a hardware-assisted cryptographic commitment construct oracle (\cco) that supplies verifiers with expensive one-time commitments without signer interaction. We realized \cco~via secure enclaves and created three \hases~instantiations: 
{\em (i) Post-Quantum \pqhases:} achieves post-quantum security and forward security simultaneously. \pqhases~transforms the one-time hash-based HORS \cite{Reyzin2002} into multiple-time without consorting with heavy sub-tree construction or secure enclaves on signers. 
{\em (ii) Lightweight \lahases:} is a conventional-secure partially aggregate signature based on Curve25519 on which the standard Ed25519 operates. \lahases~avoid running expensive EC scalar multiplication or pairing operations on signers by harnessing \cco~as the supplier of the costly EC commitments. 
{\em (iii) Hybrid \hyhases:}~combines \lahases~and \pqhases~via a novel nesting approach. This achieves partial aggregation, reinforced by a PQ-FS umbrella signature. Moreover, \hyhases~solely employs standard-compliant operations, ensuring backward compatibility. Unlike previous approaches, \hases~are the first, to the best of our knowledge, to provide a conventional-PQ hybrid signature that adheres to NIST standards without incurring heavy overhead or relying on secure hardware on signers.

\textbf{Significant Improvements over Preliminary Version and Prior Works:} A small part of this paper, specifically the \pqhases~scheme, accepted in~\cite{Yavuz:HASES:ICC2023}. Our current article makes substantial new contributions over its preliminary version on algorithmic novelty, formal security analysis, and experiment/implementation fronts (more than 18 pages of this 28‐page manuscript are new content):

\sloppy  \underline{{\em (1) Introducing \lahases~as non-interactive signer-optimal}} \underline{{ \em aggregation via \cco:}} 
Hash-based digital signatures (e.g., \pqhases) lack aggregation due to their non-algebraic structure. 
Most recent lightweight digital signatures (e.g., \cite{vallent2021efficient}) are instantiated from the primitive BLS \cite{vallent2021efficient} which lack efficient signature generation.
%
EC-based signatures (e.g., Schnorr \cite{SchnorrQ}) offer improved efficiency but still require expensive commitment generation on low-end signers, which results into a linear storage and bandwidth overhead. 
Numerous Schnorr-based aggregate signatures (e.g., \cite{chen2022half}) offer efficient verification but incur interaction between signers, thus unpractical for IoT.
Various techniques (e.g., \cite{Yavuz:2012:TISSEC:FIBAF, Yavuz:CNS:2019, nouma2023practical}) have been developed to mitigate this bottleneck. For example, FI-BAF \cite{Yavuz:2012:TISSEC:FIBAF} involves precomputing commitments during key generation and storing them on verifiers beforehand. Yet, this approach is impractical for large DT-enabled IoT networks due to its linear verifier storage overhead per user.
ESEM \cite{Yavuz:CNS:2019} separates commitment generation from signing and delegates it to a set of distributed servers. While this enhances signing speedup, it becomes susceptible to networks delays and non-colluding server assumptions. 
To the best of our knowledge, \lahases~is the first to incorporate secure enclaves for commitment generation. This allows for non-interactive and efficient aggregate signing with minimal storage overhead for verifiers. Verifiers can flexibly request one-time commitments in offline mode or on-demand. \lahases~also offers improved signing efficiency and produce aggregate signatures that are $10\times$ smaller than the initial \pqhases.

\underline{\em (2) Introducing \hyhases~via a Novel Strong Nesting Str-} \underline{\em -ategy:}
Our initial \pqhases~offers high signing efficiency and provides both post-quantum and forward security. However, it lacks aggregation which is crucial for reducing bandwidth and battery usage on low-end devices. On the other hand, \lahases~efficiently aggregates signatures and requires minimal cryptographic storage for verifiers. However, it falls short of providing post-quantum security, which is essential for long-term security in distributed DT applications. Note that both \pqhases~and \lahases~adhere to standard compliance and backward compatibility. As discussed in Section \ref{subsec:RelatedWork}, previous hybrid signatures constructions have not investigate the combinations of digital signatures with different security features (e.g., FS, aggregation). To the best of our knowledge, \hyhases~is the first hardware-assisted solution that offers signer-optimal hybrid capabilities, achieving aggregation, forward security, and PQ promise simultaneously. This is achieved by introducing a novel nesting method that combines \lahases~and \pqhases~digital signatures. 

\underline{\em (3) Expanding Performance Analysis with Further Optimiza-} \underline{\em ions:} We fully implemented all \hases~schemes on both commodity  hadware and IoT devices and tested them on actual DT datasets. We added new experiments and comparisons with the state-of-the-art PQ, aggregate, and hybrid schemes. Our experiment results showcases a speedup over initial version thanks to the use of NIST standard Ascon \cite{dobraunig2021ascon} for lightweight hashing and key derivation.


\vspace{4pt}
\textbf{Desirable Properties of \hases~Over Prior Works:}
We outline desirable properties as follows:

\underline{{ \em(1) Signer Computation and Energy Efficiency:}}  \pqhases~needs only a small-constant number of hash calls (i.e., $\approx 20$) per FS signing making it $1542 \times$ and $16.28\times$ faster than \xmssmt~and Dilithium, which is the only FS and the most signer efficient NIST PQC standard (not FS), respectively. \pqhases~is $9.34\times$ faster than Ed25519 which is neither PQ nor FS. 
\lahases~also runs only a few hash calls and modular arithmetic operations whose costs are negligible. This makes it $762\times$ and $183\times$ faster than BLS and C-RSA, respectively.  \pqhases~and \lahases~are $40\times$ and $99\times$ more energy efficient than BLISS and Ed25519 which are the only feasible PQ and conventional-secure alternatives on the 8-bit device, respectively. 
The \hyhases~is several magnitudes more efficient in signing with smaller tag sizes compared to alternative combinations (e.g., \xmssmt-C-RSA, Dilithium-BLS),  while also offering stateful signing and backward compatibility.

\underline{\em (2) Compact Signatures:} The signature size of \pqhases~is identical to \hors, making it the most compact signature among its PQ counterparts (e.g., $10\times$ smaller than only PQ-FS alternative \xmssmt). The signature size of \lahases~is comparable to that of the standard non-aggregate Ed25519 and the short signature BLS, but without expensive EC scalar multiplication and map-to-point operations, respectively. \hyhases~offers the smallest AS and FS signature with PQ security, which is $44\times$ smaller than the most compact Dilithium-BLS combination.

\underline{\textit{(3) Non-Interactive and Scalable Multi-Users:}} \sloppy  \hases~schemes lifts the burden of conveying/certifying one-time commitments from the signer, thereby making it independent from \cco~and verifiers. \cco~can supply any public commitment of a valid signer to verifiers either beforehand or on-demand, with adjustable storage overhead, permitting a scalable  management service for massive-size DT-enabled IoT networks with millions of users. 

\underline{\textit{(4) Architectural Feasibility and Cryptographic Agility:}}  (i) Unlike some hardware-based solutions (e.g., \cite{ouyang2021scb,boneh2019post}), \hases~does not require a secure enclave on the signer, which is not feasible for low-end IoT.  (ii) \hyhases~is the first, to the best of our knowledge, to offer a hybrid conventional-PQ signature with partial aggregation and FS by only using standard primitives (i.e., SHA-256, Curve25519). Thus, it can expedite the PQC transition for DTs with its backward compatibility. 


\underline{\textit{(5) High Security:}} (i) \pqhases~only relies on SHA-256, and therefore is free from rejection/Gaussian sampling that permits devastating side-channel attacks in its lattice-based counterparts (e.g., BLISS-I \cite{espitau2017side}). (ii) The stateful signing of \hases~avoids vulnerabilities of weak pseudo-random generators typically found in low-end IoT devices. (iii) \pqhases~achieve both PQ and FS properties for long-term security guarantees. (iv) The signature aggregation offers resiliency against truncation attacks due to the all-or-nothing feature. 

\underline{\textit{(6) Full-fledge Implementation}}: We fully implemented all \hases~schemes on 8-bit MCU and commodity hardware at signers. For verifier side, we used a commodity hardware equipped with Intel SGX as the \cco. Our implementation can be found at: 
~\fbox{\url{https://github.com/SaifNOUMA/HYHASES}} \\
\vspace{-2mm}

\section{Preliminaries}
\label{preliminaries}

The acrynoms and notations, used in the paper, are described in Table \ref{tab:acronyms}.

\begin{table*}[ht!]
	\caption{List of notations and acronyms}\label{tab:acronyms}
	\vspace{-2mm}
	\centering
	
	\resizebox{\textwidth}{!}{
		\begin{tabular}{|c | l | c | l  | @{}c@{} | @{}c@{} | @{}c@{} | @{}c@{} | @{}c@{} | @{}c@{} | @{}c@{} | @{}c@{} | }
			\hline
			\textbf{Notation} &  \textbf{Description} &  \textbf{Acronym} & \textbf{Description}    \\ \hline
			$\|$  &  string concatenation & IoT & Internet of Things  \\ \hline
			$|x|$ & bit length of variable $x$ & MCU & Microcontroller Unit  \\ \hline
			$x \Ra \mathcal{S}$ & random selection from a set $\mathcal{S}$ & DT & Digital Twins  \\ \hline
			$\{0,1\}^*$ &  set of binary strings of any finite length & NIST & National Institute of Standards and Technology \\ \hline
			${\{q_i\}}_{i=0}^{n-1}$ & set of items $q_i$ for $i=0,\ldots, n-1$  & PQC & Post-Quantum Cryptography  \\ \hline
			$f: \{0,1\}^* \rightarrow \{0,1\}^\kappa$  & one-way function  & PQ &  Post-Quantum security  \\ \hline 
			$\prf: \{0,1\}^* \times \{0,1\}^* \rightarrow \{0,1\}^\kappa$ & key derivation function accepts a key and a message as input & $\sk/\pk$ & Secret/Public keys  \\ \hline
			$H : \{0,1\}^* \rightarrow  \{0,1\}^{\kappa}$ & cryptographic hash function  &  FS & Forward Security \\ \hline
			$H^{(\ell)}(.)$ & $\ell$ consecutive hash evaluations  & AS & Aggregate Signature  \\ \hline
			Epoch (or Batch) $j$ & set of finite data recordings  & EC & Elliptic Curves \\ \hline
			$x_j$ & variable during epoch $j$  & PRF & Pseudo-Random Function  \\ \hline
			$x_j^\ell$ & variable during epoch $j$ in the iteration $\ell$  & ROM & Random Oracle Model  \\ \hline
			$x_j^{\ell_1, \ell_2}$ & aggregate variable during $\ell_1$ and $\ell_2$ iterations of epoch $j$  &  CCO & Commitment Construct Oracle  \\ \hline
			
		\end{tabular}
	}
	\vspace{-2.5mm}
\end{table*}

\begin{definition} \label{def:sgn}
	A signature scheme \sgn~is a tuple of three algorithms $(\kg, \ssig, \ver)$:
	\begin{enumerate}[\indent -]
		\item \underline{$(\sk, \pk) \as \sgnkg(1^\kappa)$:} Given the security level $\kappa$, it returns a private/public key pair $(\sk,\pk)$.
		\item \underline{$\sigma \as \sgnsig(\sk, M)$:} Given $\sk$ and a message $M$, the signing algorithm returns a signature $\sigma$.
		\item \underline{$b \as \sgnver(\pk, M, \sigma)$:} Given the public key $\pk$, message $M$, and its corresponding signature $\sigma$, the verification algorithm outputs a bit $b$ (if $b=1$, the signature is valid, otherwise it is invalid). 
	\end{enumerate}
\end{definition}


Aggregate signature is as Def. \ref{def:sgn} except signature generation/verification accept a set of messages.

\begin{definition} \label{def:asgn}
	\sloppy A single-signer aggregate signature scheme \asgn~is a tuple of four algorithms $(\kg, \asig, \agg, \aver)$ defined as follows:
	\begin{enumerate}[\indent -]
		\item \underline{$(\sk, \pk) \as \asgnkg(1^\kappa)$:} Given security parameter $\kappa$, it returns private/public keys $(\sk,\pk)$.
		\item \underline{$\sigma_{1,L} \as \asgnsig(\sk, \vec{M})$:} Given $\sk$ and a set of messages $\vec{M}=\{m_\ell\}_{\ell=1}^L$, it returns an aggregate signature $\sigma_{1,L}$.
		\item \underline{$\sigma_{1,L} \as \asgnagg(\{\sigma_\ell\}_{\ell=1}^L)$:} Given $L$ distinct signatures $\{\sigma_\ell\}_{\ell=1}^L$, it returns aggregate tag $\sigma_{1,L}$.
		\item \underline{$b \as \asgnaver(\pk, \vec{M}, \sigma_{1,L})$:} Given $\pk$, a set of messages $\vec{M}=\{m_\ell\}_{\ell=1}^L$, and its aggregate signature $\sigma_{1,L}$, it outputs $b$ (if $b=1$, the signature is valid, otherwise invalid). 
	\end{enumerate}
\end{definition}

Forward security \cite{ForwardSecure_MMM_02} periodically evolves the private key and deletes its previous iterations, thereby enhances breach resiliency against key compromise attacks.

\begin{definition} \label{def:fs-sgn}
	A FS signature \fsgn~have four algorithms $(\kg, \ssig, \upd, \ver)$ defined below:
	\begin{enumerate}[\indent -]
		\item \underline{$(\sk, \pk) \as \fsgnkg(1^\kappa, J)$:} Given the security parameter $\kappa$ and the maximum number of key updates $J$, it returns a private/public key pair $(\sk,\pk)$.
		\item \underline{$\sk_{j+1} \as \fsgnupd(\sk_j, J)$:} Given the private key $\sk_j$, it returns  $\sk_{j+1}$ and delete $\sk_j$, if $j<J$, else aborts. 
		\item \underline{$\sigma_j \as \fsgnsig(\sk_j, M_j)$:} Given $\sk_j$ and a message $M_j$, it returns a FS signature $\sigma_j$ as output and perform key update $\sk_{j+1} \as \fsgnupd(sk_j,J)$, if $j<J$. Otherwise, it aborts.
		\item \underline{$b \as \fsgnver(\pk, M_j, \sigma_j)$:} Given $\pk$, a message $M_j$, and its corresponding signature $\sigma_j$, the verification algorithm outputs a validation bit $b$ (if $b=1$, signature is valid, otherwise invalid). 
	\end{enumerate}
\end{definition}

%

A hybrid signature scheme consists of a combination of two (or more) distinct digital signatures. The resulting hybrid scheme is unforgeable as long as at least one of the underlying schemes remains secure. An example is the fusion of conventional and PQ signature schemes. NIST recommends hybrid schemes due to their enhanced security against unexpected algorithmic breaches and ease of transition \cite{barker2018recommendation}. Of particular interest is the nested combination strategy \cite{bindel2017transitioning}, as defined below:

\begin{definition} \label{def:hysgn}
	A hybrid signature scheme (\hysgn)~is a composition of two distinct signature schemes $\sgn_1$ and $\sgn_2$. It is described as follows:
	\begin{enumerate}[\indent -]
		\item  \underline{$(\sk, \pk) \as \hysgn.\kg(1^\kappa)$:} Given the security parameter $\kappa$, it generates $(sk_1, \pk_1) \as \sgn_1.\kg(1^\kappa)$ and $(\sk_2, \pk_2) \as \sgn_2.\kg(\kappa)$. It returns $(\sk \as \langle \sk_1,\sk_2 \rangle, \pk \as \langle \pk_1, \pk_2 \rangle )$. 
		\item  \underline{$\sigma \as \hysgnsig(\sk, M)$:} Given the private key $\sk$ and the message $M$, it computes $\sigma_1 \as \sgn_1(\sk_1, M) $ and $\sigma_2 \as \sgn_2(\sk_2, M \| \sigma_1)$. It returns $\sigma \as \langle \sigma_1, \sigma_2 \rangle$.
		\item  \sloppy \underline{$b \as \hysgnver(\pk, M, \sigma)$:}	Given $\pk$, the message $M$, and the signature $\sigma$, it computes $ b_1 \as \sgn_1.\ver(\pk_1, M, \sigma_1)$ and $ b_2 \as \sgn_2.\ver(\pk_2, M, \sigma_2)$. It returns $(b \as b_1 \wedge b_2)$.
	\end{enumerate}
\end{definition}

A hardware-assisted digital signature removes the need of supplying commitments and public keys (with certificates) from signers. It introduces a third-party entity, {\em Commitment Construct Oracle (\cco)}, equipped with a secure enclave (e.g., Intel SGX \cite{el2022benchmarking}) that issues cryptographic keys to verifiers either offline or on-demand. A PQ \cco~can be achieved by using PQ cryptographic operations within the enclave \cite{boneh2019post}. 
%
A hardware-assisted digital signature can be extended into a multi-user setting by deriving the users' private keys from a master key, which is solely stored at \cco. We call such scheme as Hardware-Assisted Mutli-User signature (\hamusgn), described in Def. \ref{def:hamus}. 

\begin{definition} 	\label{def:hamus}
	A Hardware-Assisted Multi-User signature scheme \hamusgn~consists of four algorithms $(\kg,\comconstr, \ssig,\ver)$ defined as follows:
	\begin{enumerate}[\indent -]
		\item $\underline{ (\msk, \vec{\sk}, I) \as \hamusgnkg( 1^{\kappa}, \vec{ID}, J ) }$:
		Given the security level $\kappa$, the signers' identities $\vec{ID}$~and the max number of signatures $J$, it returns a master key \msk, users' private keys $\vec{\sk}$, and the system parameters $I$.
		
		\item $\underline{ \sigma_i^j \as \hamusgnsig(\sk_i , M_i^j ) } $: 
		Given the private key $\sk_i$ of the user $ID_i$ and the message $M_i^j$, it returns the signature $\sigma_i^j$, given that the counter $j \le J$. It updates $j \as j+1$.
		
		\item $\underline{ C_i^j \as \hamusgncomconst(\msk, ID_i, j) } $: Given the master key $\msk$, the signer identity $ID_i \in \overrightarrow{ID}$, and the state $j \le J$, it returns the corresponding commitment $C_i^j$  under \msk.
		
		\item $\underline{ b_i^j \as \hamusgnver(\langle \pk_i^j, C_i^j \rangle, M_i^j, \sigma_i^j) }$: 
		Given $\pk_i^j$, the commitment $C_i^j$, a message $M_i^j$, and its associated signature $\sigma_i^j$,  it returns a bit $b_i^j$, with $b_i^j=1$ meaning {\em valid}, and $b_i^j=0$ otherwise.
	\end{enumerate}
\end{definition}

{\em Our \hamusgn~instantiations:} 
can be classified to several types of signature schemes as folllows:
\begin{enumerate}[\indent -]
	\item \sloppy {\em Aggregate-based signatures:} The signature generation and verification algorithms of the aggregate \hamusgn~(\hamuasgn) follows Def. \ref{def:asgn}. Hence, \hamuasgn~follows \AEUCMA~security model (see Def. \ref{Def:AEUCMA:HAMU}).  
	\item {\em Forward-secure signatures (see Def. \ref{def:fs-sgn}):} A commitment construct algorithm (\comconstr)~returns one-time commitments based on the private key updates at the signer. Thus, the forward-secure \hamusgn~(\hamufsgn) follows the \FEUCMA~security model (see Def. \ref{def:FEUCMA:HMU}).
	\item {\em Hybrid signatures  (see Def. \ref{def:hysgn})}:  A hybrid \hamusgn (\hyhamu) can be constructed from various \hamusgn~instantiations. In section \ref{sec:proposed_schemes}, we discuss a \hyhamu~signature, built from a fusion of an aggregate \hamuasgn~and a FS-PQ \hamufsgn~digital signatures. 
\end{enumerate}

	\section{Models}
\label{sec:models}

\subsection{System Model}
\label{system-model}

Our system model is suitable for a DT-enabled IoT network, wherein a large number of low-end devices (e.g., sensors) and various metaverse devices (e.g., camera, AR headset, smart glasses) authenticate their generated data (e.g., images, haptic biometrics, Electrocardiogram (ECG)-signals) and offload them to a remote edge server. Our model consists of three entities:

{\em (i) Signers:}  In a DT-enabled metaverse framework, various low-end devices (e.g., implants, sensors) and metaverse equipment (e.g., camera, headset) generate continuous multimedia data (e.g., haptic, image). This data requires offloading to a remote edge cloud for future analytics and decision-making. 
Thus, optimizing signautre generation and cryptographic payload is vital to prolong battery life. 
The low-end devices are also expected to operate in adversarial environments, under various attacks (e.g., key compromise \cite{Yavuz:2012:TISSEC:FIBAF}, quantum \cite{joseph2022transitioning}, truncation and delayed \cite{Yavuz:2012:TISSEC:FIBAF} attacks). 
Hence, an ideal authentication scheme should provide FS for breach resilience, PQ security against quantum attacks, and aggregation to protect against truncation attacks while supporting high transmission rates. 
As the signers are low-end devices, secure enclaves are not required at their side. Hence, we consider them broadcasting authenicated data without involving third parties or conveying public keys to verifiers. 

{\em (ii) Verifier:} encompass any untrusted entities (laptop, cloud server). They receive signatures from signers and one-time keys from a third-party supplier, either on-demand or offline. Verifiers are unconstrained by resource limitations and can possess a secure hardware. In our system model, the verifier represents the edge server which is responsible of receiving authenticated data from signers and conducting data processing for insightful analytics and smart decision making.
Therefore, we justify the use of secure enclaves on verifiers to exclusively store the user keys.
Meanwhile, our system model enables any other entity to perform verifications.  

{\em (iii) Commitment Construct Oracle (\cco):} We introduce a commitment constructor that supplies verifiers with one-time keys. Digital signatures featuring advanced security properties (e.g., FS, PQ security) incur linear bandwidth overhead in relation to transmitted data. This incur additional penalties on signers, rendering deployment nearly infeasible. This could drain significant energy, especially for the low-end signers. Moreover, It will add up a high bandwidth overhead and incur traffic congestion across public networks. The \cco~role is relieve signer from the burden of certifying one-time keys. \cco~can live on verifiers to minimize communication delays in on-demand requests. 
We implemented \cco~with Intel SGX which is widely available on cloud plateforms \cite{el2022benchmarking}. However, \hases~can be realized with {\em any} secure hardware offering a trusted execution environment.

\vspace{2mm}
\subsection{Threat and Security Model}
\label{security-model}

\sloppy We define the standard {\em Forward-secure Existential Unforgeability against Chosen Message Attack} (\FEUCMA)~\cite{ForwardSecure_MMM_02} by incorporating \cco. 
It serves as a threat model of \hamufsgn. The adversary \A~obtain $J$ FS-signatures under a challenge \pk. A \hamufsgn~scheme is proven to be \FEUCMA~based on the experiment in Def. \ref{def:FEUCMA:HMU}, wherein \A~is given four oracles defined as follows: 
%

%
%
%

\begin{enumerate}[(i)]
	\item {\em Random Oracle \ro:} \A~is given access to a random oracle from which it can request the hash of any message $M_j$ of her choice, up to $q_s'$ messages. Note that in our proofs (see Appendix B), cryptographic hash functions $H_{k=0,1,2}$ are modeled as random oracles via \ro~(Random Oracle Model (ROM)) \cite{katz2020introduction}. We note that some of our schemes explicitly use \ro, while others do not query \ro~but the security of their base signature scheme is secure in ROM (so as ours).
	\item {\em Signing oracle \Sig:} \A~is provided with a signing oracle \Sig~on a message $M_j$, of her choice. \A~can query \Sig~up to $q_s$ individual messages in total as described, until she decides to \enquote{break-in}.
	\item {\em Break-in Oracle \Breakin:} \A~is provided with break-in oracle \Breakin~which provides the current private key $\sk_j$. Formally, if \A~queried $j<J$ individual messages to \Sig, then the \Breakin~oracle returns the $(j+1)^{\text{th}}$ private key $\sk_{j+1}$ to \A. Otherwise, if $j \ge J$, then the break-in oracle rejects the query, as all private keys were used. 
	\item {\em Commitment Construct Oracle \comconstrmsk:} \A~is provided with commitment construct oracle \comconstrmsk~on a public commitment of her choice. We note that this commitment may either be a component of public key or serve as the public key itself. 
\end{enumerate}

\begin{definition} \label{def:FEUCMA:HMU}
	\FEUCMA~experiment  $\mathit{Expt}_{\hamufsgn}^{\FEUCMA}$ is as:
	
	Experiment $\mathit{Expt}_{\hamufsgn}^{\FEUCMA}(\mathcal{A})$:
	\begin{enumerate}[\indent \indent]
		\item $ (\msk, \overrightarrow{\sk}, I) \as \hamufsgnkg(1^{\kappa}) $
		\item $ (M^* , \sigma^*) \as \mathcal{A}^{\ro,~\Sig,~\comconstrmsk, \Breakin} (.) $
		\item If $ \hamufsgnver(\langle \pk, C^* \rangle, M^* , \sigma^*)=1 $ and $M^*$ was not queried to $ \Sig$ where $C^{*}$ is the  output of \comconstrmsk, return 1, else, return 0.
	\end{enumerate}
	The \FEUCMA~advantage of \A~is  $		\advhamufs = $ $Pr[\mathit{Expt}_{\hamufsgn}^{\FEUCMA}(\mathcal{A})]=1] $,where \A~have time complexity $t$, making at most $q_s'$ queries to \ro,  $q_s$ queries to $\hamufsgn.\Sig$ and $\hamufsgn.\comconstrmsk$ combined, and one query to $\Breakin$.
\end{definition}

%
%
%

We define the {\em Aggregate Existential Unforgeability Under Chosen Message Attack} (\AEUCMA)~\cite{BLS:2004:Boneh:JournalofCrypto} security model that captures the threat model of a \hamuasgn~scheme (in single-signer setting).  

\begin{definition}
	\label{Def:AEUCMA:HAMU}
	\AEUCMA~experiment  $\mathit{Expt}_{\hamuasgn}^{\AEUCMA}$ is as:
	
	Experiment $\mathit{Expt}_{\hamuasgn}^{\AEUCMA}(\mathcal{A})$:
	\begin{enumerate}[\indent \indent]
		\setlength{\parskip}{0pt}
		\setlength{\parsep}{0pt}
		
		\item $(\msk, \vec{\sk}, I)\leftarrow \hamuasgnkg(1^{\kappa})$,
		\item $(\vec{M}^{*},\sigma^{*})\leftarrow  \mathcal{A}^{\ro,~\hamuasgnsig_{\sk}(.), \comconstrmsk}(\pk)$,
		\item If $1 = \hamuasgnver(\pk^*, \vec{M}^*, \sigma) $ and $\vec{M}^*$ was not queried to $\hamuasgnasig_{\sk}(.)$, where $\pk^*$ was queried to \comconstr, return 1 else 0.
	\end{enumerate}
	
	The \AEUCMA~of \A~is as $\advhamuagg=Pr[\mathit{Expt}_{\hamuasgn}^{\AEUCMA}(\mathcal{A})=1]$ .
	
	The  \AEUCMA~advantage of \hamuasgn~is as $		\advhamuagg=\max_{\mathcal{A}}\{Adv_{\hamuasgn}^{\AEUCMA}(\mathcal{A})\}$, where the max is over \A~having time complexity $t$, with at most $q_s'$ queries to \ro~and $q_s$ queries to $\hamuasgnasig(.)$.
\end{definition}

We define the {\em Hybrid Existential Unforgeability Under Chosen Message Attack} (\HEUCMA) security model that captures the security model of a hybrid signature. We adopt the strong nesting technique as in \cite{bindel2017transitioning}. 
The hybrid scheme $\hysgn$ is constructed from two distinct signature schemes $\sgn_1$ and $\sgn_2$. Specifically, if either $\sgn_1$ or $\sgn_2$ are unforgeable in the classical (or quantum) random oracle model, then $\hysgn$ is unforgeable in the classical (or quantum) random oracle model. 

\begin{definition}
	The \HEUCMA~experiment $\mathit{Expt}_{\hysgn}^{\HEUCMA}$ is described as follows:
	\begin{enumerate}[\indent \indent]
		\item $ (\sk, \pk) \as \hysgnkg(1^\kappa) $, where $\sk=\langle \sk_1, \sk_2 \rangle$ and $\pk = \langle \pk_1, \pk_2 \rangle$,
		\item $ (M^*, \sigma^*) \as \mathcal{A}^{\ro, \sgnsig_{\sk_1}(.), \sgnsig_{\sk_2}(.) } (\pk) $, where $\sigma = \langle \sigma_1, \sigma_2 \rangle$,
		\item If $1=\texttt{SGN}_1\texttt{.Ver} (\pk^*_1, M^*, \sigma^*_1)$~, $1=\texttt{SGN}_2\texttt{.Ver} (\pk^*_2, M^* \| \sigma_1, \sigma^*_2)$, and $M^*$ was not queried to $\texttt{SGN}_1\texttt{.Sig}_\sk(.)$, and  $M^* \| \sigma_1$ was not queried to $\texttt{SGN}_2\texttt{.Sig}_\sk(.)$, return 1, otherwise return 0. 
	\end{enumerate}
	
	The \EUCMA~of \A~is defined as
	\begin{eqnarray*}
		\advhysgn= Pr[ \mathit{Expt}_{\hysgn}^{\HEUCMA} (\mathcal{A})=1 ] \\ = Pr[ \mathit{Expt}_{\sgn_1}^{\EUCMA} (\mathcal{A})=1 ] \cdot Pr[ \mathit{Expt}_{\sgn_2}^{\EUCMA} (\mathcal{A})=1 ]
	\end{eqnarray*}
	
	The \EUCMA~advantage of \sgn~is defined as
	\begin{eqnarray*}
		\advhysgn=\max_{\mathcal{A}}\{Adv_{\hysgn}^{\EUCMA}(\mathcal{A})\} \\ =\min \{ Adv_{\sgn_1}^{\EUCMA}(t, q_s, q_s')~,~ Adv_{\sgn_2}^{\EUCMA}(t, q_s, q_s') \},
	\end{eqnarray*}
\end{definition}

\textbf{Assumption 1:} The commitment construct oracle (\cco)~has a secure enclave as described in the system model. A post-quantum \cco~is easily achieved by using PQ signature primitives within the encalve.
The security of hardware-assisted signature primitives relies on trust in Intel's manufacturing process and the robustness of the SGX system. Although our focus is on implementing our proposed schemes using Intel SGX, in principle the system could be also instantiated using other isolated execution environments (e.g., Sanctum \cite{costan2016sanctum}). It is crucial to recognize the limitations of relying on trusted execution environments. For example, Intel SGX encountered various side-channel attacks (e.g., \cite{silva2022power}). Generic techniques for protection against enclave side-channel attacks are also under study in various works (e.g., \cite{lang2021informer}), therefore they are complementary to ours.

	\section{Proposed Schemes} \label{sec:proposed_schemes}

Our goal is to create a new series of cryptographic tools for data authentication and integrity featuring resilient digital signatures with advanced security and performance attributes (e.g., FS, PQ security, aggregation)  tailored for DT applications. However, designing a signature scheme that aligns with the stringent requirements of DT components poses a significant challenge \cite{el2021potential}. Consider a human DT constantly receives data from various sources (e.g., wearable sensors, implantables) to replicate the state of the real twin. 
Yet, DT-IoT components face resource constraints (e.g., battery, processing) demanding efficient signature scheme. Additionally, DTs need a PQ signature with FS to guarantee long-term assurance and breach resiliency against malware compromises.
It is also highly desirable to provide aggregation to reduce communication overhead. In the following, we propose three novel signature schemes that achieve all these seemingly conflicting performance and security goals. 

\subsection{Post-Quantum Hardware-Assisted Efficient Signatures (PQ-HASES)} \label{subsec:pq-hases}

Our initial contribution is an FS and PQ signature scheme with hardware support called \pqhases. As discussed in Section \ref{sec:Introduction}, the state-of-the-art PQ schemes are still very expensive to be deployed on low-end devices. To our knowledge, there is no open-source implementation of NIST PQC standards on highly resource-limited devices (e.g., 8-bit MCUs). It is even higher if FS is considered~\cite{yavuz2022frog}.

We created \pqhases~to address these limitations. It primarily uses a hash-based one-time signa-ture scheme (\hors) \cite{Reyzin2002}, which is also the basis for NIST's recommendation \xmssmt \cite{cooper2020recommendation} and standard SPHINCS+ \cite{bernstein2019sphincs+}. \pqhases~achieves FS by updating the private key through a hash chain. The main bottleneck for hash-based schemes stems from an expensive computation and transmission of public keys. Thus, we introduce \cco~as public-key issuer which enables a non-interactive signer that only broadcasts authenticated data. The algorithmic intuition of \pqhases~is outlined below.

\vspace{1mm}
We give the detailed algorithmic description of \pqhases~in Fig. \ref{alg:pqhases}. 

In \pqhaseskg, for a given set of users $\vec{ID}$, it first generates the master key \msk~and the \cco-related keys for key certification purposes (Step 1). It also accepts as input $(J_1,J_2)$, the number of precomputed and interleaved keys, respectively. $J$ represents the number of signatures to be generated  (Step 2). According to $J_1$, it generates the precomputed keys $\vec{\sk}_p^n$ from $\sk_1^n$ (Step 5-7). Each private seed $\sk_1^n$ is sent to its corresponding signer $ID_n$ (Step 8), while $(\msk, \vec{\sk}_p)$ are placed on the secure enclave of \cco~(Step 1).

\begin{figure*}[ht!]
	\centering
	
	\begin{minipage}{\textwidth}
		\centering
		\noindent \fbox{\parbox{\textwidth} {
				\scriptsize
				
				\vspace{-2mm}
				\begin{multicols}{2}
					\begin{algorithmic}[1]
						\Statex   $\underline{(\msk, \vec{\sk}_1, \vec{\sk}_p, I)\as \pqhaseskg(1^{\kappa},
							\vec{ID} = \{ID_n\}_{n=1}^N, J_1,  J_2)}$:
						\vspace{3pt}
						\State Generate master key $\msk \Ra \{0,1\}^\kappa$ and \cco-related keys $(\csk, \cpk) \as \sgnkg(1^\kappa)$.  Securely store $(\msk, \csk)$ at \cco.
						\State Set $J \as J_1 \cdot J_2$ and $I \as (l,k,t, J_1, J_2, J)$ 
						\For{$n = 1, \ldots, N$}
						\State $\sk_1^n \as \prf(\msk , ID_n)$
						\For{$j_1 = 2, \ldots, J_1 - 1$}
						\State $\sk_{j_1 \cdot J_2 + 1 }^n \as H^{(J_2)}( \sk_{ (j_1 - 1) \cdot J_2 + 1}^n )$
						\EndFor
						\State $\vec{\sk}_p^n \as \{ \sk_{j_1 \cdot J_2 + 1 }^n  \}_{j_1=1}^{J_1 - 1}$
						\EndFor
						\State $\vec{\sk}_1 \as \{\sk_1^n \}_{n=1}^{N}$, where each $\sk_1^n$ is provisioned to $ID_n$, as initial private key.
						\State $\vec{\sk}_p \as \{  \vec{\sk}_p^n \}_{n=1}^{N}$, are provisioned to \cco, as precomputed private keys.
						\State \Return ($\msk, \vec{\sk}_1, \vec{\sk}_p,  I$)
					\end{algorithmic}
					\algrule
					
					\begin{algorithmic}[1]
						\Statex   $\underline{(C_j, \sigma_{C_j}) \as \pqhasescomconstr(\msk, \csk, \vec{\sk}_p, ID, j)}$: Require $ID\in \vec{ID}$ and $j \le J$
						\vspace{3pt}
						\State $j_1 \as \lfloor \frac{j-1}{J_2} \rfloor $ and $j_2 \as (j-1) \mod J_2$
						\State Load $\sk_{j_1 \cdot J_2+1}$ from $\vec{\sk}_p$ and derive $\sk_j \as H^{(j_2)} ( \sk_{j_1 \cdot J_2 + 1 } )$
						\For {$\ell=1, \ldots, t$}
						\State $v_\ell \as H(s_\ell)$, where $s_\ell \as \prf(\sk_j, \ell)$
						\EndFor
						\State $C_j \as (v_1, v_2, \ldots, v_t)$
						\State $\sigma_{C_j} \as \sgnsig(\csk, v_1 \| \ldots \| v_t)$
						\State \Return $(C_j , \sigma_{C_j})$
					\end{algorithmic}
					\algrule
					
					\begin{algorithmic}[1]
						\Statex $\underline{\sigma_j \as \pqhasessig(\sk_j, M_j)}$: 
						Init $(j=1)$, require $ j \le J$
						\vspace{1pt}
						\State $h \as H(M_j)$ and split $h$ into $k$ substrings $\{h_\ell\}_{\ell=1}^k$ such that $|h_\ell|=\log_2{t}$, where  each $h_\ell$ is interpreted as an integer $x_\ell$
						\For{$\ell = 1, \ldots, k$}
						\State $s_{\ell} \as \prf(\sk_j, x_{\ell})$
						\EndFor
						\State Update $\sk_{j+1} \as H(\sk_j) $, delete $\sk_j$, and increment $j \as j+1$
						\State	\Return $\sigma_j \as (s_1, s_2, \ldots, s_k, j, ID)$
					\end{algorithmic}
					\algrule
					\begin{algorithmic}[1]
						\Statex $\underline{b\as \pqhasesver(\langle \cpk, C_j, \sigma_{C_j} \rangle, M_j,\sigma_j)}$:  Steps 1-2 can run in offline mode. 
						\vspace{3pt}
						\State $(C_j, \sigma_{C_j}) \as \pqhasescomconstr(\msk, \csk, \vec{\sk}_p, ID, j)$ 
						\State \textbf{if} $\sgnver(\cpk, v_1 \| \ldots \| v_t, \sigma_{C_j})=1$ \textbf{then continue else return} $0$
						\State Execute Step (1) in \pqhasessig
						\If{$H(s_{\ell}) = v_{x_{\ell}}, \forall \ell \in [1, k]$ and $1 \leq j \leq J$} \Return $b=1$ \textbf{else} \Return $b=0$ 		\EndIf
					\end{algorithmic}
				\end{multicols}
				\vspace*{-2mm}
		}}
	\end{minipage}
	\caption{ The proposed PQ hardware-assisted digital signature with forward-security (\pqhases) }
	\label{alg:pqhases}
\end{figure*}

\begin{figure}
	\centering
	\includegraphics[width=90mm]{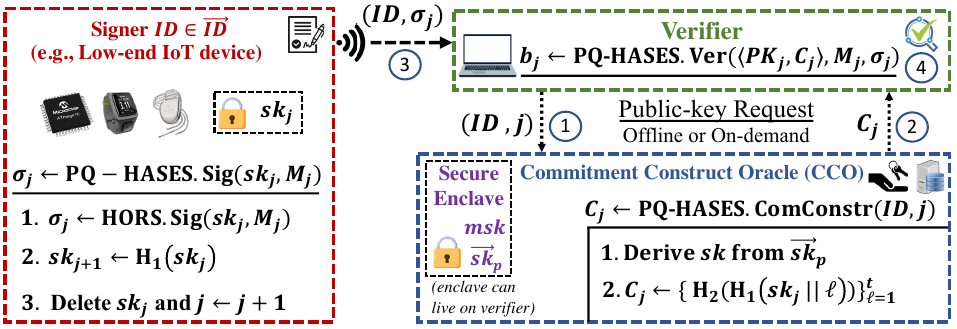}
	\caption[Caption for LOF]{High-level description of the \pqhases~scheme}
	\label{fig:pq-hases}
\end{figure}

\pqhasescomconstr~harnesses \cco~to offer a trustworthy and flexible public commitment supply and identity management service for verifiers, without requiring interaction with signers. The verifier requests the one-time commitment $C_j$ of $ID$ for the state $j$. \cco~first identify the corresponding pre-computed key (Step 1) and then derive the $j^{\text{th}}$ secret key (Step 2). Finally, it generates the commitment $C_j$ and returns it to the verifier (Step 3-5). \cco~can derive any one-time commitment $C_j$ of any $ID \in \vec{ID}$ on demand, making \pqhases~fully scalable for millions of users with an adjustable $\mathcal{O}(J_1)$ cryptographic data storage. Moreover, the verifier can obtain any public key(s) $1 \le j \le J$ from \cco~in batches before receiving signatures (\pqhasescomconstr~is independent from signer). This permits verifiers to immediately verify signatures. Also, \cco~can either present on the verifier machine (e.g., laptop), or a nearby edge-cloud server and therefore can effectively deliver public commitments on demand. In order to authenticate the commitments, we used a generic digital signature and generate a signature on the public commitment (Step 6).

The signing algorithm \pqhasessig~relies on \hors~signature but with a FS pseudo-random number generation. Given $\sk_j$, the signer computes subset resilient indexes and derives the \hors~one-time signature components (steps 1-4 in Fig. \ref{alg:pqhases}). The current private key is updated and the previous key is deleted. \hasessig~offer near-optimal signing efficiency by running a single \hors~call and a single hash for the update. The signer do not use or interact with secure hardware.

The algorithm \pqhasesver~also depends on \hors, leveraging \pqhasescomconstr~to acquire certified public commitments which are verified via the \cco~public key (Step 1-2). Upon having a valid commitment which represent the public key, it continues by performing the \hors~signature verification algorithm (Step 3-4). It is important to note that \pqhases~distinct itself from symmetric-key methods (e.g., MACs alone or use of secure hardware to compute/verify MACs) by providing public verifiability and non-repudiation. The verifier can validate the signatures with an offline interaction with \cco, which only supplies commitments and does not perform the verification.

\vspace{2mm}
{\em Optimizations:} Our design allows verifiers to request the one-time public keys before signature verification, which avoids the network and \cco~computation delays. \pqhases~offers an adjustable storage-computation trade-off by controlling the number of pre-computed keys stored at \cco. We observed that \cco~could benefit from an optimizer that manage the data storage overhead for each signer. For instance, this optimizer could be implemented with a Reinforcement Learning algorithm that learns from previous verifier requests~\cite{mao2016resource} in order to reduce the computation delay.

\subsection{Lightweight Aggregate Hardware-Assisted Efficient Signatures (LA-HASES)} \label{subsec:agg-hases}

In DT-enabled settings, extensive network traffic arises from end-to-end interactions between physical objects (e.g., humans, sensors) and their digital replicas (e.g., cloud server). Therefore, aggregation becomes pivotal in curbing network delays, enhancing service quality, conserving transmission energy in wireless setups, and consequently prolonging battery life. Moreover, it alleviates local storage on signers, thus creating more room for the main core-specific DT applications.

\begin{figure*}[ht!]
	\centering
	\begin{minipage}{\textwidth}
		\centering
		\noindent \fbox{\parbox{\textwidth} {
				\scriptsize
				
				\begin{multicols}{2}
					\begin{algorithmic}[1]
						\Statex   $\underline{(\msk,  \vec{\sk}, \vec{\pk}, I)\as \lahaseskg(1^{\kappa}, \vec{ID} = \{ID_n\}_{n=1}^N, J, L)}$:
						\vspace{3pt}
						\State Generate large primes $q$ and $p \ge q$ such that $q|(p-1)$. Select a generator $\alpha$ of the subgroup G of order q in $\mathbb{Z}_q^*$. Set $I \as (p,q,\alpha, J, L, St: (ID_n~,~j=1) ), ~\forall n \in [1,N]$
						\State Generate master key $\msk \Ra \{0,1\}^\kappa$ and \cco-related keys $(\csk, \cpk) \as \sgnkg(1^\kappa)$.  Securely store $(\msk, \csk)$ at \cco.
						\For{$n = 1, \ldots, N$}
						\State $y_n \as \prf (\msk , ID_n) \mod q$, and $Y_n \as \alpha^{y_n} \mod p$
						\EndFor
						\State $\vec{\sk} \as \{ \csk, y_1, \ldots, y_N \}$, where $y_n$ is provisioned to $ID_n$.
						\State $\vec{\pk} \as \{ \cpk, Y_1, \ldots, Y_N \} $
						\State \Return $(\msk, \vec{\sk}, \vec{\pk}, I)$
					\end{algorithmic}
					\algrule
					
					\begin{algorithmic}[1]
						\Statex $\underline{\delta_{1,u} \as \lahasesagg( \{ \delta_\ell \in \sigma \}_{\ell=1}^u )}$: 
						\vspace{1pt}
						\State \textbf{if} $\delta \in \mathbb{Z}_q^*$ \textbf{then} $\delta_{1,u} \as \sum_{\ell=1}^{u} \delta_\ell \mod q$
						\State	\Return $\delta_{1,u}$
					\end{algorithmic}
					\algrule
					
					\begin{algorithmic}[1]
						\Statex   $\underline{(C_{j}^{1,L}, \sigma_{C_j}) \as \lahasescomconstr(\msk, \csk,~ ID, j)}$: Require $ID \in \vec{ID}$ and $j \le J$
						\vspace{3pt}
						\State $y \as \prf (\msk , ID) \mod q$, and $r_j \as \prf(y , j \| 2) \mod q$
						\State $R_j^{1,L} \as \alpha^{\sum_{\ell=1}^{L} r_j^{\ell} \mod q}  \mod p$~, where $r_j^{\ell} \as \prf(r_j , \ell) \mod q, \forall \ell \in [1,L]$
						\State $\sigma_{C_j} \as \sgnsig(\csk, C_j^{1,L})$, where $C_j^{1,L} \as R_j^{1,L} $
						\State \Return $(C_j^{1,L}, \sigma_{C_j})$
					\end{algorithmic}
					
					\begin{algorithmic}[1]
						\Statex $\underline{\sigma_j^{1,L} \as \lahasessig(y, \vec{M}_j)}$: require $j \le J$ and $\vec{M}_j=(m_j^1, \ldots, m_j^L)$
						\vspace{1pt}
						\State $x_j \as \prf(y , j || 1)~$, and $r_j \as \prf(y , j \| 2)$
						\State $s_{j}^{1,0} \as 0$
						\For{$\ell=1, \ldots, L$}
						\State $x_j^\ell \as \prf(x_j , \ell)$, and $r_j^\ell \as \prf(r_j , \ell) \mod q$
						\State $s_j^\ell \as r_j^\ell - e_j^\ell \cdot y \mod q$, where $e_j^\ell \as H( m_j^\ell \| x_j^\ell )$
						\State $ s_j^\ell \as \lahasesagg(s_j^{1,\ell-1}, s_j^{\ell}) $
						\EndFor
						\State Update  $St: (ID~,~j \as j+1)$ 
						\State	\Return $\sigma_j^{1,L} \as \langle s_j^{1,L}, x_j, St \rangle$
					\end{algorithmic}
					\algrule
					
					\begin{algorithmic}[1]
						\Statex $\underline{b \as \lahasesver(\langle \cpk, \pk, C_j^{1,L} \rangle, \vec{M}_{j}, \sigma_j^{1,L})}$: require $\vec{M}_j=\{ m_j^{\ell} \}_{\ell=1}^L$. 
						Steps 1 can be run offline. 
						\vspace{3pt}
						\State $(C_j^{1,L}, \sigma_{C_j}) \as \pqhasescomconstr(\msk, \csk, ID, j)$, where $C_j^{1,L} = R_j^{1,L} $ (offline mode or on-demand)
						\State \textbf{if} $\sgnver(\cpk, C_j^{1,L}, \sigma_{C_j})=1$ \textbf{then continue else return} $0$
						\State Set $e_j^{1,0} \as 0$ 
						\For{$\ell=1, \ldots, L$}
						\State $x_j^{\ell} \as \prf(x_i^j , \ell)$
						\State $e_j^{1,\ell} \as e_j^{1,\ell-1} + e_j^{\ell} \mod q$, where $ e_j^{\ell} \as H (m_j^{\ell} \| x_j^{\ell} ) $
						\EndFor
						\If{$R_j^{1,L} = Y^{e_j^{1,L}} \cdot \alpha^{s_j^{1,L}} $} \Return $b=1$	\textbf{else} \Return $b=0$ \EndIf
					\end{algorithmic}
					
				\end{multicols}
		}}
	\end{minipage}
	\caption{The proposed lightweight aggregate-based hardware-assisted digital signature (\lahases)}
	\label{alg:lahases}
	\vspace*{-2mm}
\end{figure*}

\begin{figure}
	\centering
	\includegraphics[width=90mm]{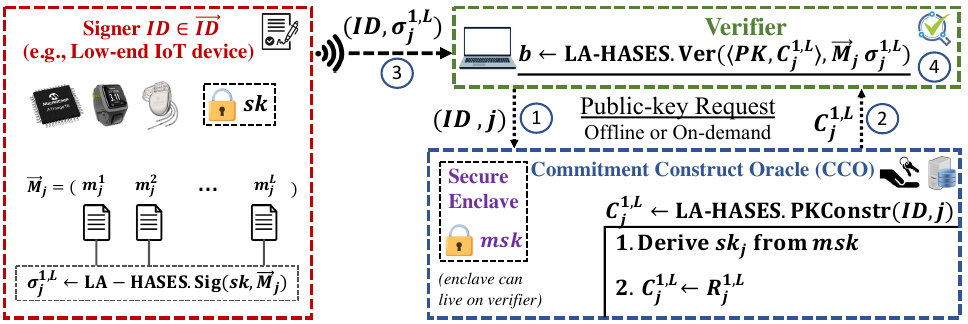}
	\caption[Caption for LOF]{High-level description of the \lahases~scheme}
	\label{fig:la-hases}
\end{figure}

{\em Algorithmic Novelty and Differences with Previous Works:}
To the best of our knowledge, no aggregate PQ signature exists that is computation and bandwidth efficient. For example, a recent lattice-based AS scheme \cite{boneh2020one} offer logarithmic compression w.r.t number of signatures. However, it is limited to either one-time non-interactive or multiple-time interactive setting. Another approach converts conventional aggregate signatures into the lattice domain via Fiat-Shamir with Aborts \cite{boudgoust2023sequential}. Yet, this incur higher computational overhead and minimal compression gains. 

Among conventional-secure alternatives, recent IoT authentication protocols (e.g., \cite{vallent2021efficient, li2020permissioned}) predominantly rely on the primitive pairing-based AS scheme, BLS \cite{BLS:2004:Boneh:JournalofCrypto}. As outlined in Section \ref{subsec:RelatedWork}, BLS impose a burdensome signing overhead on resource-constrained devices, primarily due to costly map-to-point and modular exponentiation operations it entails. Furthermore, these works typically lack performance assessments on low-end MCUs, leaving their practicality unjustified.

EC-based AS schemes (e.g., \cite{Yavuz:2012:TISSEC:FIBAF, nouma2023practical, Yavuz:CNS:2019}) offer a balance between compact signature size and efficient signing when compared to above alternatives. However, it comes at the cost of linear communication and storage overhead for one-time commitments. 
In fact, FI-BAF \cite{Yavuz:2012:TISSEC:FIBAF} tranforms Schnorr \cite{SchnorrQ} into an an aggregate digital signature by replacing $H(M \| R)$ with $H(M \| x)$, where $x$ is a one-time random key. This allows detaching the commitment computation from signature generation and performing it during key generation in offline mode. Note that the main bottleneck in signature generation of Schnorr-based schemes lies in calculating the commitment $R$ which entails expensive modular exponentiation per signing. This is why FI-BAF offer a significantly more cost-effective signature generation compared to Schnorr.
Nevertheless, verifiers are required to store commitments, linear w.r.t number of messages. This makes it not scalable for our specific DT applications.

\lahases~aims to optimize FI-BAF \cite{Yavuz:2012:TISSEC:FIBAF}, which represents the improved version of the primitive Schnorr \cite{SchnorrQ}, upon which recent lightweight authentication protocols (e.g., \cite{vallent2021efficient}) depends. 
Aligned with \pqhases, we create a Lightweight Aggregate HArdware-Assisted Efficient Signature (\lahases) that uses \cco~to transmit costly EC commitments to verifiers, enabling a non-interactive efficient aggregate signing. Therefore, \lahases~achieve near-optimal signing with few modular arithmetic operations and hash calls while also leaving verifers with  low and flexible storage overhead. That is, verifiers can request commitments in batches in offline mode or on-demand. Unlike FI-BAF, \cco~supplies verifiers with constant-size aggregate commitments per batch of input messages which significantly lower the communication and storage overhead on verifier side. Moreover, \cco~can live on verifiers, thereby enabling zero communication overhead.

We give a detailed algorithmic description of \lahases~digital signature scheme in Fig. \ref{alg:lahases}.

In \lahaseskg, it generates system parameters, which include the EC-based settings, the maximum number of generated  signatures $J$, and the batch size $L$ (Step 1). For a given set of users $\vec{ID}$, it generates the master key $\msk$ and \cco~related keys (Step 2) (provisioned to \cco), derives the users' private keys $\vec{sk}$ and provision each private key $\sk_n$ to its corresponding user $ID_n$ (Step 3-5).

\lahasescomconstr~offers a secure commitment supply service for the verifiers. In offline or on-demand mode, the verifier can request the public commitments of any user $ID_n$ and for any given state $j$. \cco~ identifies the user's private key and computes  $r_j$ of the requested epoch $j$ (Step 1). Then, it constructs the aggregate public commitment $R_j^{1,L}$ (Step 2), and returns it as  output (Step 3).

\lahasessig~computes a fixed-length signature for a set of $L$ messages. 
First, it computes the seed $x_j$ and commitment $r_j$ for epoch $j$ (Step 1). The aggregate signature $s_j^{1,L}$ is computed without expensive EC scalar multiplications, relying on few \prf~and hash calls, and arithmetic operations per item (Step 3-6). 
Finally, the signer state is updated and the signature is returned (Steps 7-8).

\lahasesver~can receive the certified public commitments from \cco~either in offline or on-demand mode (Step 1). It computes the one-time ephemeral keys $\{e_j^\ell\}_{\ell=1}^{L}$ to recover the aggregate key $e_j^{1,L}$ (Step 3-6). Finally, it checks the verification equation and outputs a validation bit (Step 7). 


\subsection{HYbrid Hardware-ASsisted Efficient Signatures (HY-HASES)}
\label{subsec:hyhases}

The migration from conventional-secure standards to PQC is expected to be a significant effort, posing challenges due to the heterogeneous nature of current software/hardware platforms. Given the substantial expenses and risks associated with PQC (e.g., algorithmic breaks),   NIST  recommends hybrid architectures as a transitional strategy, that fuses conventional and PQ schemes to leverage their distinct attributes~\cite{barker2018recommendation}.  Hybrid signatures promote cryptographic agility which is crucial to resist single points of failures \cite{ott2019identifying}. It offers a better resiliency to emerging hardware technology and algorithmic breakthroughs like recently broken NIST PQC candidates. Moreover, the hybrid solution should ideally be standard-compliant for ease of transition. Our main idea is to leverage conventional-secure aggregate \lahases~to enhance signing efficiency, complemented with \pqhases~for per-batch FS and PQ securities. This results in our proposed \hyhases~digital signature. It is a combination of \lahases~and \pqhases, enriched by a strong novel nesting technique.

\begin{figure*}[ht!]
	\centering
	\includegraphics[width=160mm]{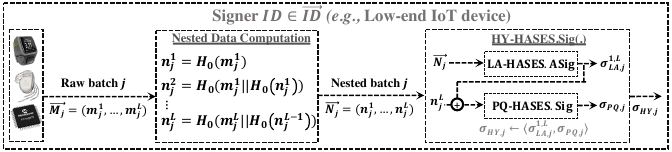}
	\caption[Caption for LOF]{High-level description of \hyhases~signature generation algorithm}
	\label{fig:hy-hases}
\end{figure*}

\begin{figure*}[ht!]
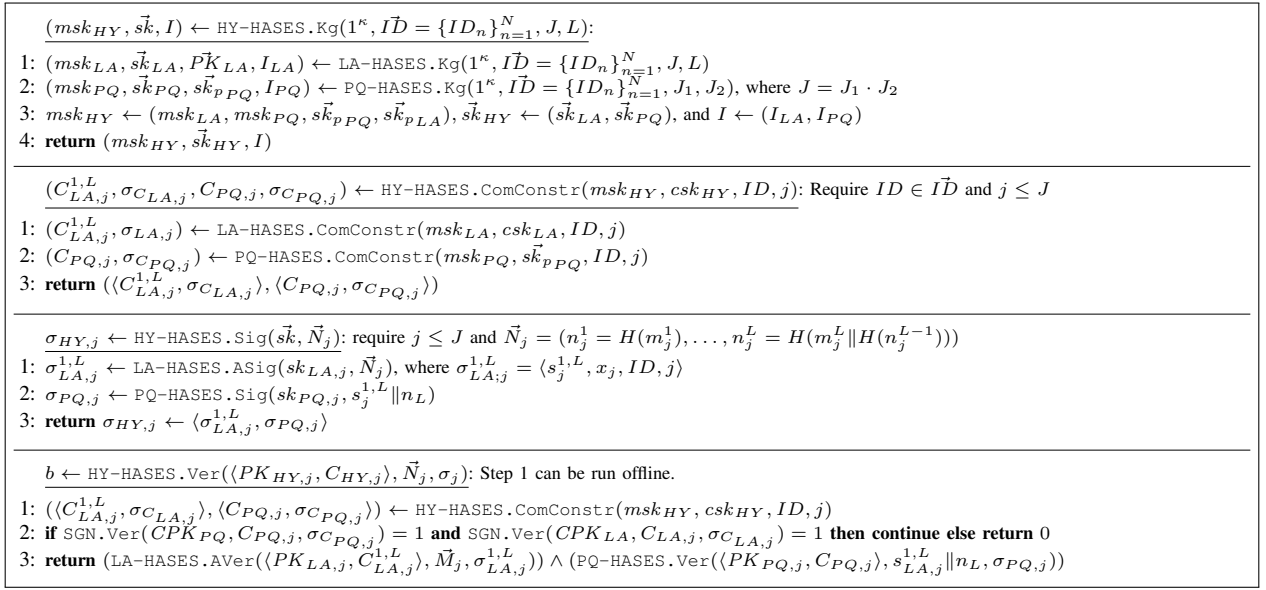

	\centering
	\begin{minipage}{0.96\textwidth}
		\centering
		\noindent \fbox{\parbox{0.93\textwidth} {
				\scriptsize
				\begin{algorithmic}[1]
					\Statex   $\underline{(\msk_{HY},  \vec{\sk}, I)\as \hyhaseskg(1^{\kappa}, \vec{ID} = \{ID_n\}_{n=1}^N, J, L)}$:
					\vspace{3pt}
					\State $ (\msk_{LA}, \vec{\sk}_{LA}, \vec{\pk}_{LA}, I_{LA} ) \as \lahaseskg(1^\kappa, \vec{ID} = \{ ID_n \}_{n=1}^N, J, L ) $
					\State $ ( \msk_{PQ}, \vec{sk}_{PQ}, \vec{\sk_p}_{PQ}, I_{PQ} ) \as \pqhaseskg(1^\kappa, \vec{ID} = \{ ID_n \}_{n=1}^N, J_1, J_2 ) $, where $J = J_1 \cdot J_2$
					\State $ \msk_{HY} \as ( \msk_{LA}, \msk_{PQ}, \vec{sk_p}_{PQ}, \vec{sk_p}_{LA}), \vec{\sk}_{HY} \as (\vec{\sk}_{LA}, \vec{\sk}_{PQ} ) $, and $ I \as (I_{LA}, I_{PQ}) $
					\State \Return $(\msk_{HY}, \vec{\sk}_{HY}, I)$
				\end{algorithmic}
				\algrule
				\begin{algorithmic}[1]
					\Statex   $\underline{ (C_{LA,j}^{1,L}, \sigma_{C_{LA,j}} , C_{PQ,j}, \sigma_{C_{PQ,j}} ) \as \hyhasescomconstr(\msk_{HY}, \csk_{HY}, ID, j)}$: Require $ID \in \vec{ID}$ and $j \le J$
					\vspace{3pt}
					\State $ (C_{LA, j}^{1,L}, \sigma_{LA,j}) \as \lahasescomconstr(\msk_{LA}, \csk_{LA}, ID, j) $
					\State $ (C_{PQ,j}, \sigma_{C_{PQ,j}}) \as \pqhasescomconstr(\msk_{PQ}, \vec{\sk_p}_{PQ}, ID, j) $
					
					\State \Return $(\langle C_{LA,j}^{1,L}, \sigma_{C_{LA,j}} \rangle, \langle C_{PQ,j}, \sigma_{C_{PQ,j}} \rangle) $
				\end{algorithmic}
				\algrule
				\begin{algorithmic}[1]
					\Statex $\underline{\sigma_{HY,j} \as \hyhasessig(\vec{\sk}, \vec{N}_j)}$: require $j \le J$ and $\vec{N}_j=(n_j^1=H(m_j^1), \ldots, n_j^{L}= H( m_j^{L} \| H(n_j^{L-1})))$
					\vspace{1pt}
					\State $ \sigma_{LA,j}^{1,L} \as \lahasessig(\sk_{LA,j}, \vec{N}_j) $, where $\sigma_{LA;j}^{1,L} = \langle s_j^{1,L}, x_j, ID, j \rangle$
					
					\State $ \sigma_{PQ,j} \as \pqhasessig(\sk_{PQ,j}, s_j^{1,L} \| n_L) $
					
					\State	\Return $\sigma_{HY,j} \as \langle \sigma_{LA,j}^{1,L}, \sigma_{PQ,j} \rangle$
				\end{algorithmic}
				\algrule
				\begin{algorithmic}[1]
					\Statex $\underline{b \as \hyhasesver(\langle  \pk_{HY,j}, C_{HY,j} \rangle , \vec{N}_j, \sigma_j)}$:  Step 1 can be run offline. 
					\vspace{3pt}
					\State $(\langle C_{LA,j}^{1,L}, \sigma_{C_{LA,j}} \rangle, \langle C_{PQ,j}, \sigma_{C_{PQ,j}} \rangle) \as \hyhasescomconstr(\msk_{HY}, \csk_{HY}, ID, j)$
					\State \textbf{if} $\sgnver(\cpk_{PQ}, C_{PQ,j},\sigma_{C_{PQ,j}})=1$ \textbf{and} $\sgnver(\cpk_{LA},C_{LA,j},\sigma_{C_{LA,j}})=1$ \textbf{then continue else return} $0$
					\State \Return $ (\lahasesver(\langle \pk_{LA,j}, C_{LA,j}^{1,L} \rangle, \vec{M}_j, \sigma_{LA,j}^{1,L} ) ) \wedge ( \pqhasesver(\langle \pk_{PQ,j}, C_{PQ,j} \rangle, s_{LA,j}^{1,L} \| n_L, \sigma_{PQ,j} ) ) $
				\end{algorithmic}
		}}
	\end{minipage}
	\caption{The proposed hybrid hardware-assisted digital signature (\hyhases)}
	\label{alg:hyhases}
\end{figure*}


{\em Algorithmic Novelty and Differences with Previous Works:} As discussed in Section \ref{subsec:RelatedWork}, recent hybrid authentication schemes fail to address practicality issues on resource-constrained networks. For example, in a benchmark study \cite{paul2020towards}, a variety of hybrid digital signatures were tested, using  both standard-compliant conventional (e.g., ECDSA) and PQ (e.g., Dilithium) digital signatures. However, our findings demonstrate that deploying standard digital signatures on low-end MCUs is impractical due to their high signature generation costs. 
Furthermore, standard digital signatures lack essential security features, specifically aggregation and forward security, which are crucial for optimizing network throughput and safeguarding against physical attacks.

Hybrid digital signatures involve merging various digital signatures through a nesting algorithm to achieve security of all underlying signature schemes. However, existing combination methods (e.g., \cite{bindel2017transitioning}) fail to handle merging digital signatures with varying security properties. To our knowledge, no prior work has tackled the integration of an AS scheme with an FS and PQ digital signature. 
Nonetheless, we recognize the importance of simultaneously achieving these security features while maintaining efficient signature generation, especially for low-end IoT devices. In this context, we present a novel nesting strategy that uniquely blend \lahases~and \pqhases~signature schemes to construct the hybrid \hyhases, achieving all desirable properties.

We give a detailed algorithmic overview of \hyhases~in Fig. \ref{alg:hyhases} and high-level depiction in Fig. \ref{fig:hy-hases}.
We adopt the aggregate \lahases~for intra-batch input messages, in order to minimize both the cryptographic payload and the energy use during continuous data offload. To achieve the PQ and FS features, we utilize \pqhases~for inter-batches. \hyhases~key generation involve both \lahaseskg~and \pqhaseskg~algorithms. Given set of users $\vec{ID}$, \hyhaseskg~initially generates master key and system parameters for \pqhases~and \lahases~(Step 1-2). The master secret key of \hyhases~is composed of $\msk_{LA}$ and $\msk_{PQ}$. The private key vector $\vec{\sk}$ and the system parameters $I$ are also compromised of $LA$ and $PQ$ components.

Combiners enable the construction of hybrid signatures. For example, strong nesting (see Def. \ref{def:hysgn}) involves one scheme generating a signature on an input message, which is then passed alongside the message as input to the second scheme. Yet, they are not applicable to merging the underlying \lahases~and \pqhases~schemes.
Indeed, it is inefficient to upload a data batch (i.e., $\mathcal{O}(L)$) to the FS and PQ scheme (\pqhases) as input due to the $\mathcal{O}(L)$ linear signature overhead.

To counter this, we construct a nested vector $\vec{N}_{j}$  sequentially from the raw input vector $\vec{M}_j=\{m_{j}^\ell\}_{\ell=1}^L$. This envolves concatenating the current message $m_j^{\ell}$ with the hash of the previous nested element $H_0 ( n_{j}^{\ell-1} )$ and hashing the resulting string (see Fig. \ref{fig:hy-hases} and \hyhasessig~in Fig. \ref{alg:hyhases}). The elements of $\vec{N}_j$ are: $n_{j}^1=H_0(m_j^1)$ and  $n_j^{\ell}=H_0(m_j^{\ell} \| H_0(n_j^{\ell-1} ) )~,~\forall \ell=2,\ldots, L$. The final element $n_j^{L}$ represents a {\em holistic digest} of the data batch $\vec{M}_j$. 
In \hyhasessig, we generate the aggregate signature $\sigma_{LA,j}^{1,L}$ (Step 1). Next we compute the PQ signature by inputting concatenation of the aggregate signature and the last element of nested data (i.e., $s_{j}^{1,L} \| n_{j,L}$) to the \pqhasessig~scheme. 
The resulting hybrid signature $\sigma_{HY,j}^{1,L}$ consists of the concatenation of both signatures $\sigma_{LA,j}^{1,L}$ and $\sigma_{PQ,j}$.
For signature verification via \hyhasesver, it receives one-time commitments for both schemes (Step 1). The hybrid signature is verified only if both of the \lahases~and \pqhases~verification algorithms are valid (Step 2).

	\section{Security Analysis} \label{sec:SecurityAnalysis}


We formally prove that \pqhases, \lahases, and \hyhases~are \FEUCMA, \AEUCMA, and \HEUCMA~secure (in the random oracle model \cite{katz2020introduction}) in Theorem \ref{the:PQ-HASES-SecurityTheorem}, \ref{the:LA-HASES-SecurityTheorem}, and \ref{the:HY-HASES-SecurityTheorem}, respectively. We ignore the terms that are negligible in terms of $\kappa$. The full proofs are given in Appendix \ref{appendixB}.
Our claimed security proofs follows the seminal works (e.g., \cite{katz2020introduction}). Hence, it inherits similar limitations related to unsufficient security model or unpractical assumptions \cite{wang2014anonymous}.

\begin{theorem} \label{the:PQ-HASES-SecurityTheorem}
	If a polynomial-time adversary \A~can break the \FEUCMA~secure \pqhases~in time $t $ and after  $ q_s  $ signature and commitment queries, and $q_s'$ queries to \ro, with a break-in query, then one can build polynomial-time algorithm \F~that breaks the EU-CMA  secure \hors~in time $ t' $ and $ q_{s }' $ queries under Assumption 1. 
	\begin{eqnarray*}
		\advpqhases & \le & J \cdot \advhors
	\end{eqnarray*}
	$\text{,~where } q'_s = q_s+1 \text{ ~and~ } \mathcal{O}(t')=\mathcal{O}(t) +  k \cdot H_0$
\end{theorem}

\begin{theorem} \label{the:LA-HASES-SecurityTheorem}
	If a polynomial-time adversary \A~can break the \AEUCMA~secure \lahases~in time $t$ and after $q_s$ signature and public key queries, and $q_s'$ queries to the random oracle \ro, the one can build polynomial-time algorithm \F~that breaks the \dlp~problem in time $t'$ and $q_s'$  queries under Assumption 1. 
	\begin{eqnarray*}
		\advlahases  \le  \advdll~, \text{~where } t' = \mathcal{O}(t) + \mathcal{O}(\kappa^3)
	\end{eqnarray*}
\end{theorem}

\begin{theorem} \label{the:HY-HASES-SecurityTheorem}
	If either \lahases~or \pqhases~is unforgeable in the classical (or quantum) random oracle model, then the hyrbid scheme \hyhases~is unforgeable in the classical (or quantum) oracle model, respectively, under Assumption 1. 
	\begin{eqnarray*}
		\advhyhases =\min \{ Adv_{\pqhases}^{\FEUCMA}(t, q_s, q_s',1)~ \\ ,~ Adv_{\lahases}^{\AEUCMA}(t, q_s, q_s') \}
	\end{eqnarray*}
\end{theorem}

	\section{Performance Analysis} \label{sec:performance_analysis}

This section examines the performance of \hases~schemes and contrasts them with their counterparts.

\subsection{Evaluation Metrics and Experimental Setup}
\label{subsec:metrics}

\noindent \underline{\em Evaluation metrics:}
We compare \hases~schemes and their counterparts based on:
(i) signer's efficiency (i.e., key sizes, computation, energy use)
(ii) advanced security aspects (e.g., FS, PQ security),
(iii) compliance with standards and ease of transition, 
(iv) verifier's computational overhead.

\vspace*{1mm}
\noindent \underline{\emph{Selection Rationale of Compared Counterparts}}:  We selected counterparts to represent the performance of primary conventional-secure aggregate and PQ digital signature families.  
Specifically, we consider the following conventional aggregate signatures: 
{\em (i) Pairing-based:} BLS \cite{BLS:2004:Boneh:JournalofCrypto} is based on pairing maps. It is well-suited for multi-user scenarios, characterized by compact signature and public key sizes. 
{\em (ii) Factorization-based:} C-RSA \cite{Yavuz:TDSC:OutsourcedDB} providing mainly a fast signature verification. 

For PQ signatures, we consider:
{\em (i) Lattice-based:} Dilithium  \cite{ducas2018crystals} is a NIST PQC standard, featur-ing efficient signing with moderate key sizes compared to other PQC candidates. BLISS \cite{espitau2017side} is the sole PQ signature with an open-source implementation for low-end devices (i.e., 8-bit AVR MCU). 
{\em (ii) Hash-based:} SPHINCS+ \cite{bernstein2019sphincs+} is the sole hash-based NIST PQC standard. Also, \xmssmt~\cite{cooper2020recommendation} is a FS signature, serving as both an RFC standard and a NIST recommendation.

We assess the performance of \hyhases~by comparing it to a nested combination of the most relevant conventional aggregate and PQ signatures from our listed counterparts.  We note that numerous orthogonal efforts targeted IoT authentication as discussed in Section \ref{subsec:RelatedWork}. Many of these works use digital signatures as a fundamental component, thus potentially benefitting from our schemes. Therefore, our focus remains on comparing \hases~to combinations of above digital signatures.

\vspace*{1mm}
\noindent \underline{\em Parameter Selection:} 
We set the security parameter as $\kappa=128$ and SHA-256 as our cryptographic hash function $H$. We used Ascon as our \prf~for key derivation calls. Ascon have been selected as the NIST standard for lightweight cryptography since it offer high efficiency and security measures. We used the Curve25519 \cite{Ed25519} (as  NIST's  FIPS 186-5 standard, 256-bit public keys) for our EC-based \lahases~scheme to offer standard compliance and a better computational efficiency. 
We set the number of signers as $N=2^{20}$ and the maximum number of messages to be signed as $J=2^{20}$ (as in XMSS \cite{XMSS}). 
The user identity list $\vec{ID}=\{ID_i\}_{i=1}^N$ are considered as MAC addresses. 
In \pqhases, we choose $I_{PQ} \as \{ l=256, t=1024, k=16 \} $ for a security level $\kappa=128$. Our choice prioritizes an optimal signer efficiency as it only requires a few ($\approx 20$) hash calls to perform one signature generation. The resource-limited signers are non-interactive and do not communicate public keys, and therefore the parameter $t$ does not impact the signing performance. 
The composite modulo size in C-RSA is $|n|=2048$.
The generic digital signature $\sgn$ that is used in the commitment construction algorithm is instantiated with the NIST PQC standard Dilithium \cite{ducas2018crystals}.

\subsection{Performance Evaluation and Comparison}
\label{subsec:performance_analysis}

\subsubsection{Performance on Commodity Hardware}
We now outline the evaluation of signer and verifier sides using the commodity hardware.

\underline{\em Hardware and Software Configuration:} 
\hases~schemes were fully implemented on the signer and verifier sides using commodity hardware. 
We used a desktop with an Intel i9-9900K@3.6 GHz processor and 64 GB of RAM. Our approach is based on OpenSSL\footnote{\url{https://github.com/openssl/openssl}} and Intel SGX SSL\footnote{\url{https://github.com/intel/intel-sgx-ssl}} open-source libraries. 
The benchmarking schemes utilized a dataset\footnote{\url{https://ieee-dataport.org/documents/mechanocardiograms-ecg-reference}} with samples from accelerometers, gyroscopes, and ECG samples collected from 29 volunteers, simulating a DT-enabled health monitoring application. This involves secure offloading of data to a cloud for disease detection. The average message size is 32 bytes, with minimal extra overhead for larger messages (due to hash compression).

\begin{figure*}[ht!]
	\centering
	
	\begin{subfigure}[b]{0.32\textwidth}
		\centering
		\includegraphics[scale=0.15]{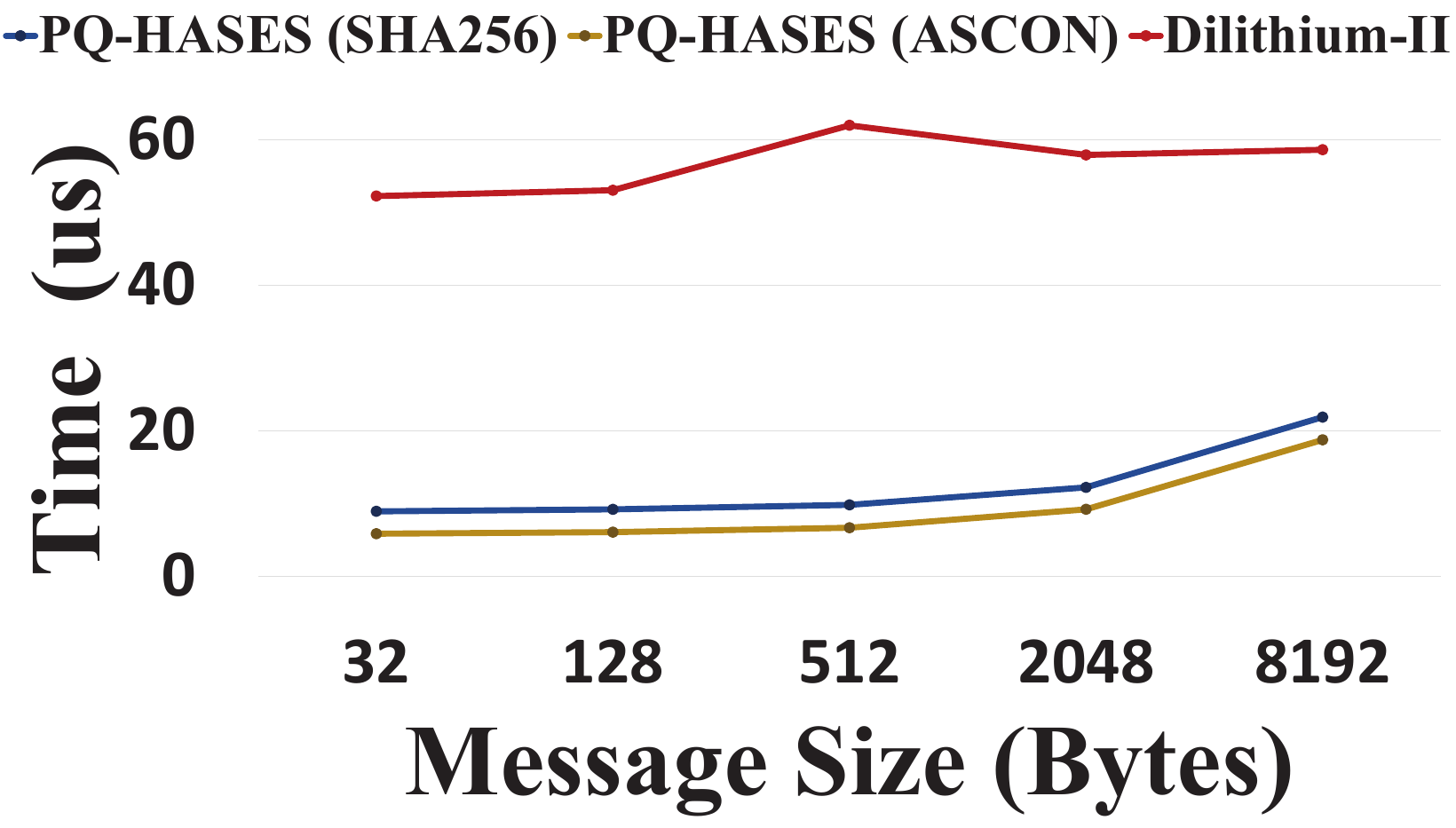}
		\label{fig:pqhases_msg}
	\end{subfigure}
	\hfill
	\begin{subfigure}[b]{0.32\textwidth}
		\centering
		\includegraphics[scale=0.15]{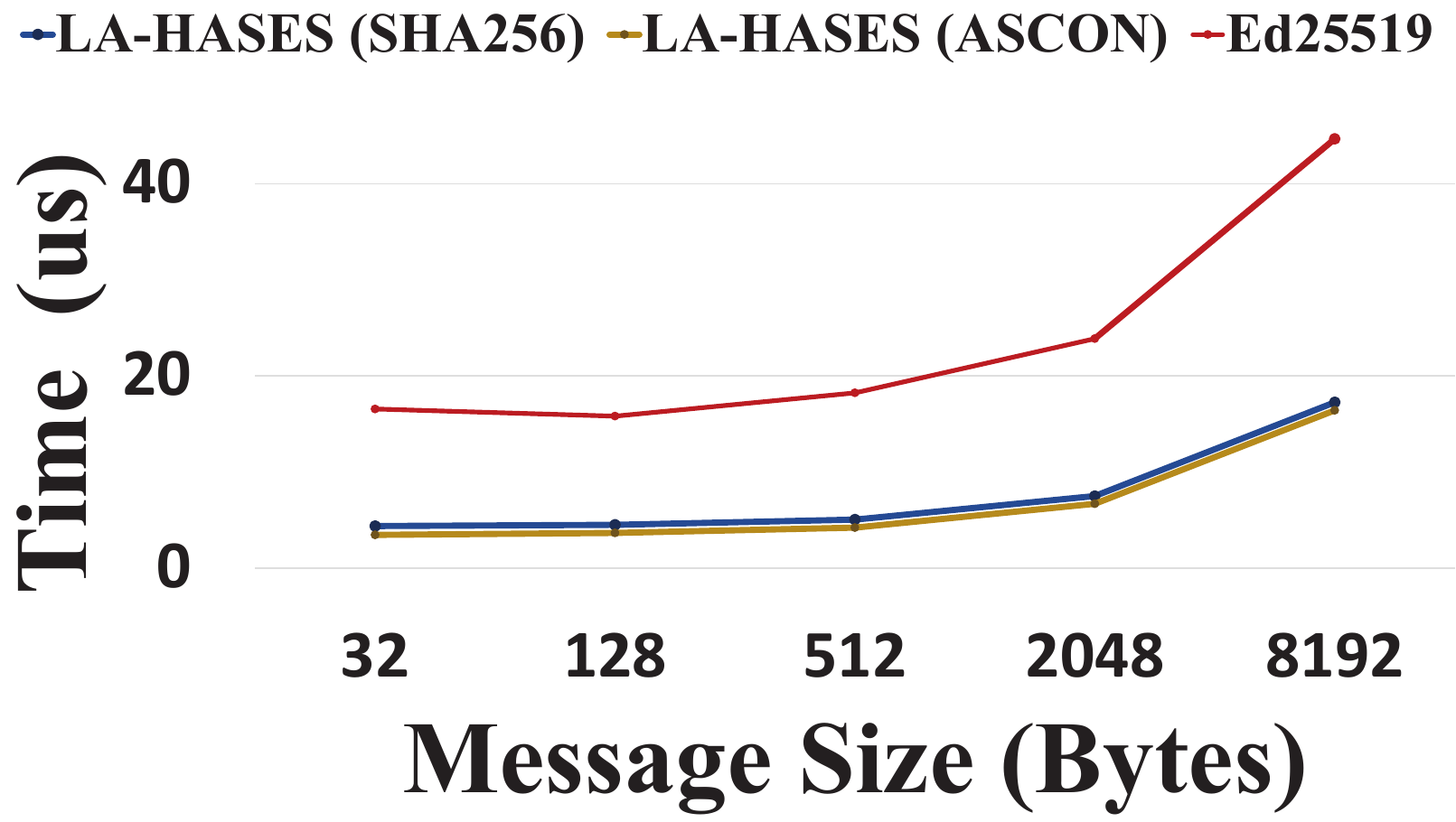}
		\label{fig:lahases_msg}
	\end{subfigure}
	\hfill
	\begin{subfigure}[b]{0.32\textwidth}
		\centering
		\includegraphics[scale=0.15]{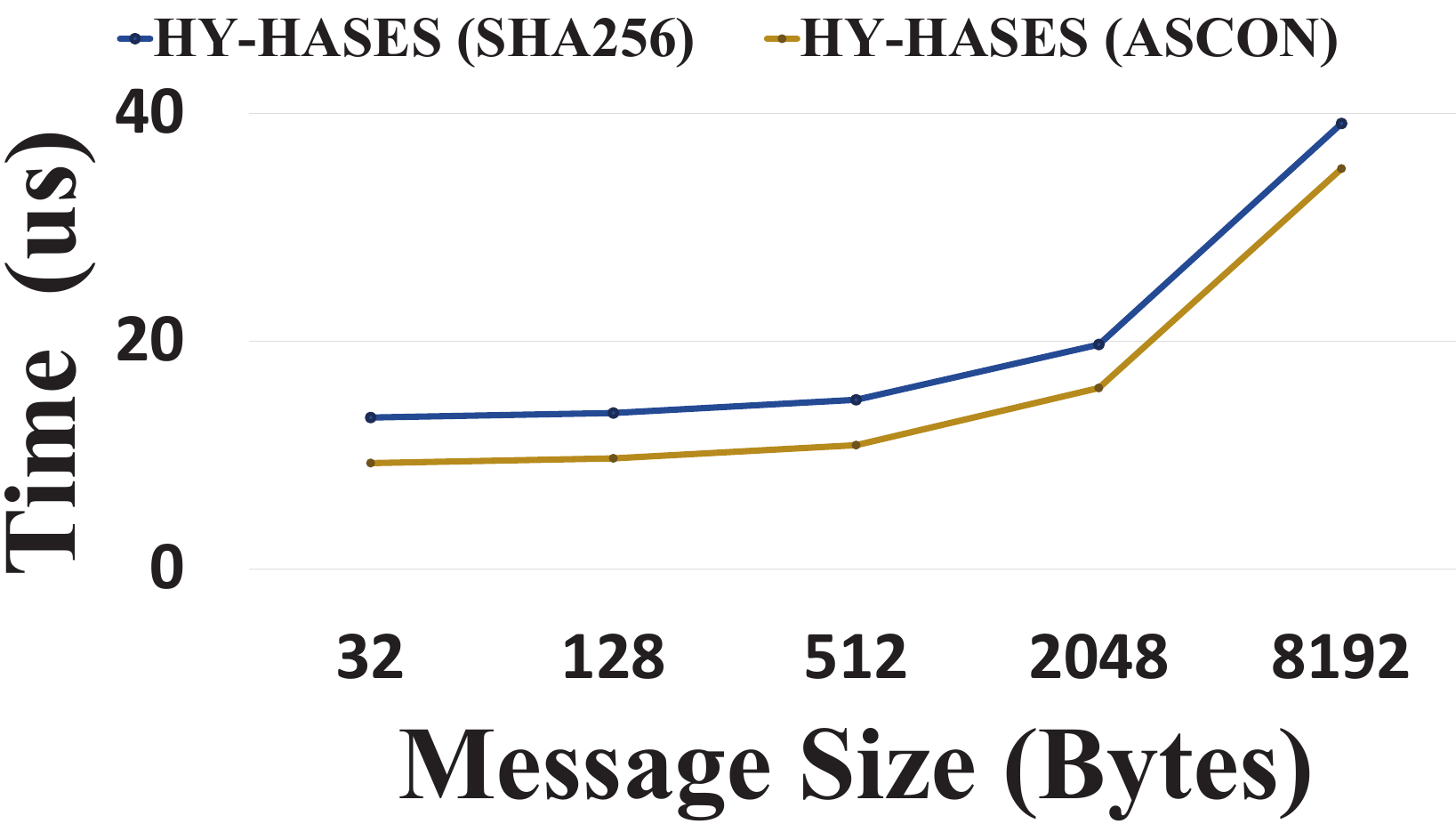}
		\label{fig:hyhases_msg}
	\end{subfigure}
	
	\caption{Signature generation of \hases~schemes and their counterparts on Commodity Hardware}
	\label{fig:hases_msg}
\end{figure*}

\underline{\em Performance Analysis:} Table \ref{tab:hysas-commodity} illustrates the overall performance of \hases~schemes and their counterparts at the signer and verifier sides. We present the main takeaways as follows:

$\bullet$ {\em Signature Generation:} The conventional-secure aggregate \lahases~scheme offers a speedup of $762\times$ and $183\times$ over the pairing-based BLS and factorization-based C-RSA schemes, respectively. This speedup is attributed to \lahases's distinct commitment separation via \cco. In contrast, BLS and C-RSA necessitate resource-intensive map-to-point and scalar multiplication operations, respectively, both scaling linearly with batch size. 
\pqhases~also outperforms its counterparts with $1542\times$ speedup over its only FS and PQ hash-based counterpart, \xmssmt~\cite{cooper2020recommendation}. It is also $16.28\times$ faster than the non-forward-secure NIST PQC standard, Dilithium-\romannum{2} \cite{ducas2018crystals}. \pqhases~achieves this while executing only a few hash and PRF calls ($\approx 20$), ensuring robust security and fast signing. Additionally, it exhibits a $74\times$ signing speedup over BLISS-\romannum{1} which lacks NIST PQC standardization and remains vulnerable to devastating side-channel and timing attacks \cite{espitau2017side}.
We further investigate the signing capabilities of \hases~schemes by examining the signature generation in relation to variations in the input message size. As illustrated in Fig. \ref{fig:hases_msg}, \pqhases~surpasses its most efficient counterpart, Dilithium \cite{ducas2018crystals}, by several orders of magnitude. We conducted benchmark tests on \hases~schemes using two different hash instantiations: the NIST standard SHA256 and the NIST lightweight PRF function Ascon. We observe that opting for Ascon as the PRF choice results in relatively higher speedup compared to \lahases. This is attributed to the frequent hash and PRF calls for the hash-based HASES. We conducted a comparison between \lahases~and the standard Ed25519, demonstrating the performance advantages offered by \lahases.

$\bullet$  {\em Signature Size:} \lahases~provides a compact signature, $4\times$ smaller than RSA-2048 and equal in size to Ed25519. The signature size of \pqhases~is $195\times$ and $16\times$ smaller than the forward-secure \xmssmt~and the NIST PQC standard Dilithium, respectively.

$\bullet$ {\em Signature Verification and Verifier Storage:} The delay of the \cco~computation is parameterized by $\Delta$ in \pqhases~whereas in the aggregate \lahases~scheme it is constant. Verifiers can request one-time commitments in either {\em Offline} or {\em Online} mode. In offline mode, the \cco~delay does not affect signature verification, resulting in significant gains compared to our counterparts. For example, if the commitment's retrieval is offline, \pqhases~verification is $11\times$ and $622\times$ much faster than Dilithium-\romannum{2} (non-forward-secure) and \xmssmt~(forward-secure), respectively.  
The commitment construction delay is determined by the number of precomputed \pqhases~private keys (i.e., $J_1$) stored at \cco. 
Fig. \ref{fig:energy-consumption}-(a) depicts the variation of this \cco~storage on the \comconstr~computation delay. For minimal storage of $320$ MB in the secure enclave within \cco, the delay is the highest at $\Delta=63.64$ msec, making the offline mode the recommended option. In contrast, if only $2^{10}$ precomputed keys are stored for each user in the network, the \cco~storage increases to $32$ GB, resulting in a significant improvement in the \comconstr~delay with $\Delta=7.13$ msec during online requests. We note that $32$ GB represents only $6.25\%$ of the protected memory in the second generation of SGX, which is widely integrated into new Intel Ice Lake processors \cite{el2022benchmarking}.

\begin{table*}[ht!]
	\centering
	\caption{Performance comparison of \hases~variants and its counterparts on a commodity hardware} \label{tab:hysas-commodity}
	
	\resizebox{\textwidth}{!}{
		\begin{tabular}{|l || @{}c@{} | @{}c@{} | @{}c@{} | @{}c@{} | @{}c@{} | @{}c@{} | @{}c@{}| @{}c@{} | @{}c@{} | @{}c@{} | @{}c@{} | @{}c@{} | @{}c@{} | }
			
			\hline
			\textbf{Scheme} &  \specialcell[]{ \textbf{Signing} \\ \textbf{Time ($\mu$s)} } &  \specialcell[]{ \textbf{Priv} \\ \textbf{Key} } &  \specialcell[]{ \textbf{Sig}  \\ \textbf{Size} } &  \specialcell[]{ \textbf{Pub} \\ \textbf{Key} } &  \multicolumn{2}{c|}{\specialcell[]{ $\textbf{Ver}^{\ddagger}$ \\ \textbf{Time ($\mu$s)} } } &  \specialcell[]{ $\textbf{Verifier}^{\star}$ \\ \textbf{Storage (GB)} }  &  \specialcell[]{ \textbf{Long-Term} \\ \textbf{Security} } & \specialcell[]{ \textbf{Backward} \\ \textbf{Compatible} }  & \specialcell[]{ \textbf{Ease of} \\ \textbf{Transition} } &  \specialcell[]{ \textbf{FSec} }  &  \specialcell[]{ \textbf{Agg} \\ \textbf{Sig} }  \\ \hline \hline
			
			Ed25519~\cite{Ed25519} & $31.85$ & $0.03$  & $0.06$ & $0.03$  &  \multicolumn{2}{c|}{$78.99$}  &  $0.09$  & \tikz\pic{sema=white/0/}; & $\checkmark$ & $\times$ & $\times$ & $\times$ \\ \hline
			RSA-2048~\cite{Yavuz:TDSC:OutsourcedDB} & $413.36$ & $0.5$  & $0.25$ & $0.5$  &  \multicolumn{2}{c|}{$21.84$}  &  $0.75$  & \tikz\pic{sema=white/0/}; & $\checkmark$ & $\times$ & $\times$ & $\checkmark$ \\ \hline
			BLS~\cite{BLS:2004:Boneh:JournalofCrypto} & $1,722.62$ & $0.06$  & $0.03$ & $0.06$  &  \multicolumn{2}{c|}{$3,568.17$}  &  $0.09$  & \tikz\pic{sema=white/0/}; & $\times$ & $\times$ & $\times$ &  $\checkmark$ \\ \hline 
			LA-HASES & $\boldsymbol{2.26}$ & $\boldsymbol{0.03}$ & $\boldsymbol{0.05}$  & $\boldsymbol{0.03}$  & \multicolumn{2}{c|}{$\boldsymbol{431.4}$} & $\boldsymbol{32}$ \textbf{B} & \tikz\pic{sema=white/0/}; & $\times$ & $\times$ & $\times$ & $\checkmark$  \\ \hline \hline
			BLISS-\romannum{1}~\cite{espitau2017side} & $252.35$ & $2.00$  & $5.60$ & $7.00$  & \multicolumn{2}{c|}{$25.02$} &  $12.6$ &  \tikz\pic{sema=black/0/}; & $\times$ & $\times$ &  $\times$ & $\times$ \\ \hline
			Dilithium-\romannum{2}~\cite{ducas2018crystals} & $55.51$ & $2.53$  & $2.36$ & $1.28$  & \multicolumn{2}{c|}{$18.65$}  & $3.64$ & \tikz\pic{sema=black/0/}; & $\times$ & $\times$ & $\times$ &  $\times$ \\ \hline
			SPHINCS+~\cite{bernstein2019sphincs+} & $4,712.23$ & $0.1$  & $35.66$ & $0.05$  & \multicolumn{2}{c|}{$331.4$}  & $35.71$ & \tikz\pic{sema=black/0/}; & $\times$  & $\times$ & $\times$ &  $\times$ \\ \hline
			\xmssmt~\cite{cooper2020recommendation} & $5,213.93$ & $5.86$  & $4.85$ & $0.06$  & \multicolumn{2}{c|}{$1,057.63$} &  $4.91$ &  \tikz\pic{sema=black/0/}; & $\times$ & $\times$ & $\checkmark$ & $\times$ \\ \hline \hline
			PQ-HASES & $\boldsymbol{3.41}$ & $\boldsymbol{0.03}$ & $\boldsymbol{0.5}$   & $\boldsymbol{32}$& \multicolumn{2}{c|}{$\boldsymbol{1.7+\Delta}$} & $\boldsymbol{32}~\textbf{B} +\boldsymbol{D}$  & \tikz\pic{sema=black/0/}; & $\checkmark$ & $\times$ & $\checkmark$  & $\times$ \\ \hline \hline
			\multicolumn{13}{|c|}{ \textbf{Hybrid Aggregate Signature Constructions (numbers are for $L$ items)} } \\ \hline \hline
			
			RSA \hfill ($L\times$) + Dilithium & $424,882.39$ & $6.53$ & $2.45$   & $4.48$ &  \multicolumn{2}{c|}{$476.71$} & $6.93$ & \tikz\pic{sema=white/180/black}; & $\times$ & $\checkmark$ & $\times$ &  $\checkmark$ \\ \hline
			
			RSA \hfill ($L\times$) + \xmssmt & $430,040.81$ & $7.11$ & $2.64$   & $4.11$ & \multicolumn{2}{c|}{$5635.13$} & $6.75$  & \tikz\pic{sema=white/180/black}; & $\times$ & $\checkmark$ & $\checkmark$ & $\checkmark$ \\ \hline
			
			BLS \hfill ($L\times$) + Dilithium & $120,650.65$ & $2.78$ & $4.42$   & $7.73$ & \multicolumn{2}{c|}{$122,155.19$} & $12.15$  & \tikz\pic{sema=white/180/black}; & $\times$ & $\checkmark$ & $\times$ &  $\checkmark$ \\ \hline
			
			BLS \hfill ($L\times$) + \xmssmt & $120,650.65$ & $3.26$ & $4.61$  & $7.36$  & \multicolumn{2}{c|}{$127,313.61$} & $11.97$  & \tikz\pic{sema=white/180/black}; & $\times$ & $\checkmark$ & $\checkmark$ &  $\checkmark$ \\ \hline \hline
			
			HY-HASES \hfill ($L\times$) & $\boldsymbol{1100.04}$ & $\boldsymbol{1.88}$ & $\boldsymbol{0.55}$  & $\boldsymbol{374.02} $  & \multicolumn{2}{c|}{$\boldsymbol{644.83+\Delta}$} & $\boldsymbol{64}~\textbf{B} +\boldsymbol{D}$  & \tikz\pic{sema=white/180/black}; & $\checkmark$ &  $\checkmark$ & $\checkmark$ &  $\checkmark$ \\ \hline
		\end{tabular}
	}
	
	\begin{tablenotes}[flushleft] \scriptsize{
			\item The private/public  key and signature sizes are in KB. 
			We benchmarked the XMSSMT-SHA$20/2\_256$ variant which allows for $2^{20}$ messages to be signed. For SPHINCS+ parameters, $n=16, h = 66, d = 22, b = 6,  k = 33, w = 16 \text{ and } \kappa=128$. The batch size of input messages is $L=1024$. We set $N=2^{20}$ signers and $J=2^{10}$ as signing capability per user. $L\times$ refers to the batch size, fed to the AS schemes. 
			\tikz\pic{sema=white/0/}; and \tikz\pic{sema=black/0/}; denotes conventional and post-quantum security guarantees, respectively. \statcirc[white]{black} denotes PQ-conventional hybrid security.
			BLISS and Dilithium incurs sampling operations which may be costly for low-end devices and prone to side-channel attacks \cite{espitau2017side}.
			\item $\star$ Our system model assumes that the edge cloud performs signature verification to further analyze the multimedia data. In such case, \cco~lives on verifiers. We consider that the verifier storage for \hases~schemes consists of \cco~private keys. The verifier storage for our counterparts is the cryptographic keys (i.e., public key and its certificate) for all users in the network (i.e., $N=2^{20}$). If \cco~does not live on verifier, we assume the verifier stores public keys or obtains them along one-time commitments from \cco, online or offline.
			\item $\ddagger$ $\Delta$ denotes the \cco~delay of commitment construction in \pqhases. $\Delta$ depends on the storage of precomputed \pqhases~private keys by \cco, which is defined by $D$. Fig. \ref{fig:energy-consumption}-(a) depicts the impact of \cco~storage on the \comconstr~computation overhead.
		}
	\end{tablenotes}
\end{table*}

\subsubsection{Performance on 8-bit AVR Microcontroller}
Table \ref{tab:hases_8bit} compare the signer performance of \hases~schemes with their counterparts on the 8-bit AVR microcontroller. 

\begin{figure*}[ht!]
	\centering
	\begin{subfigure}[b]{0.29\textwidth}
		\centering
		\includegraphics[scale=0.16]{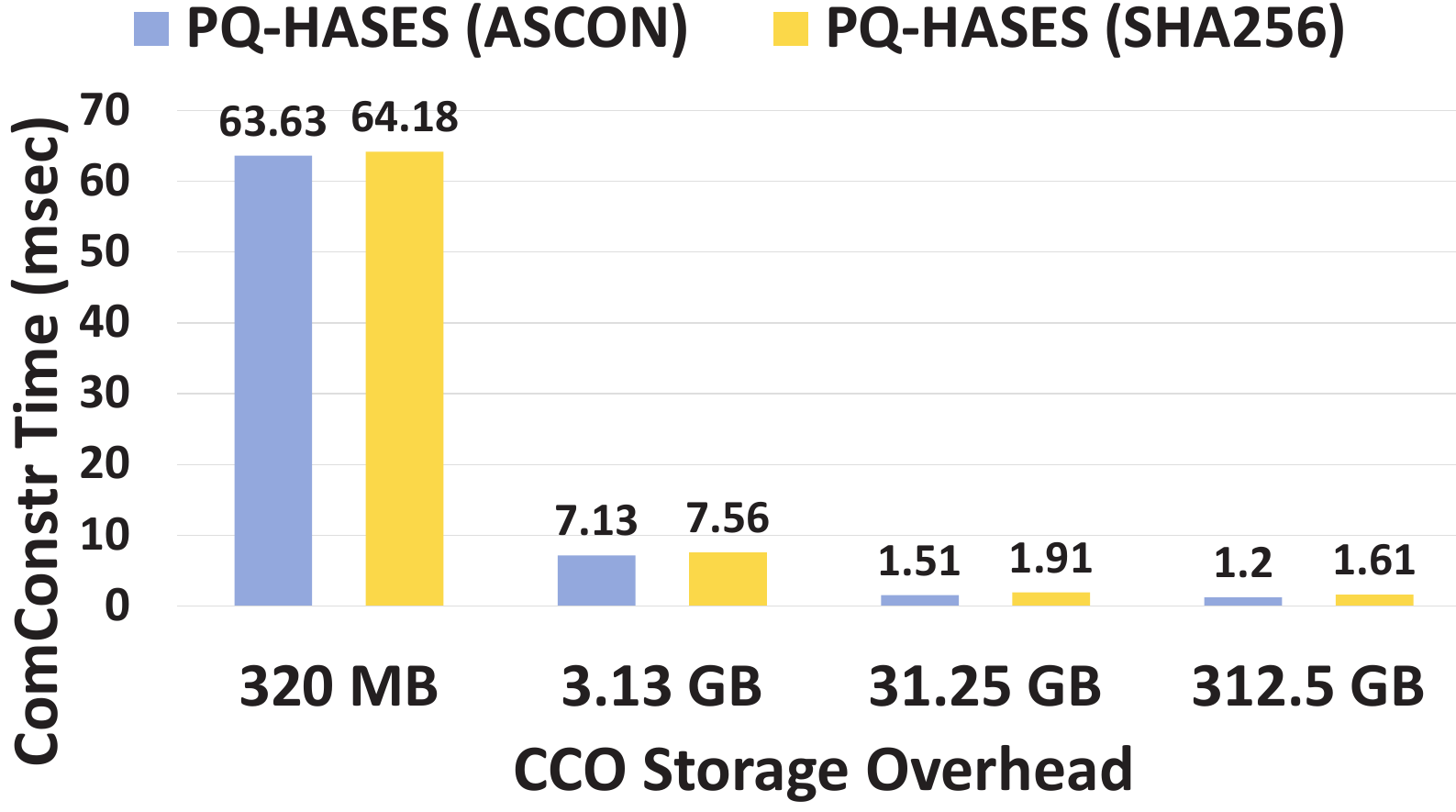}
		\label{fig:comconstr}
		\caption{\cco~Storage and Delay}
	\end{subfigure}
	\hfill
	\begin{subfigure}[b]{0.29\textwidth}
		\centering
		\includegraphics[scale=0.13]{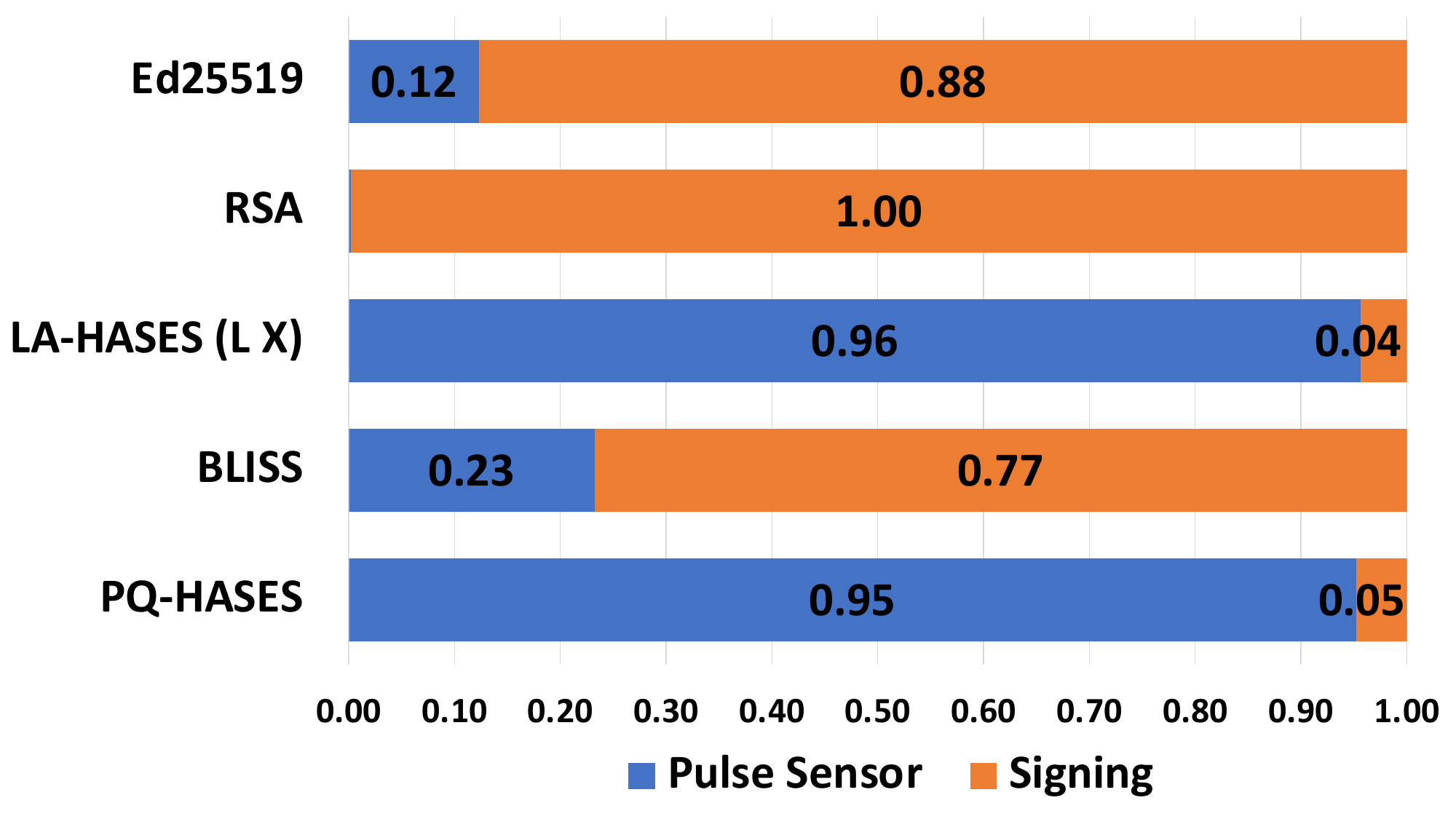}
		\label{fig:pulse_sensor}
		\caption{Energy usage}
	\end{subfigure}
	\hfill
	\begin{subfigure}[b]{0.27\textwidth}
		\centering
		\includegraphics[width=0.9\textwidth, height=0.1\textheight]{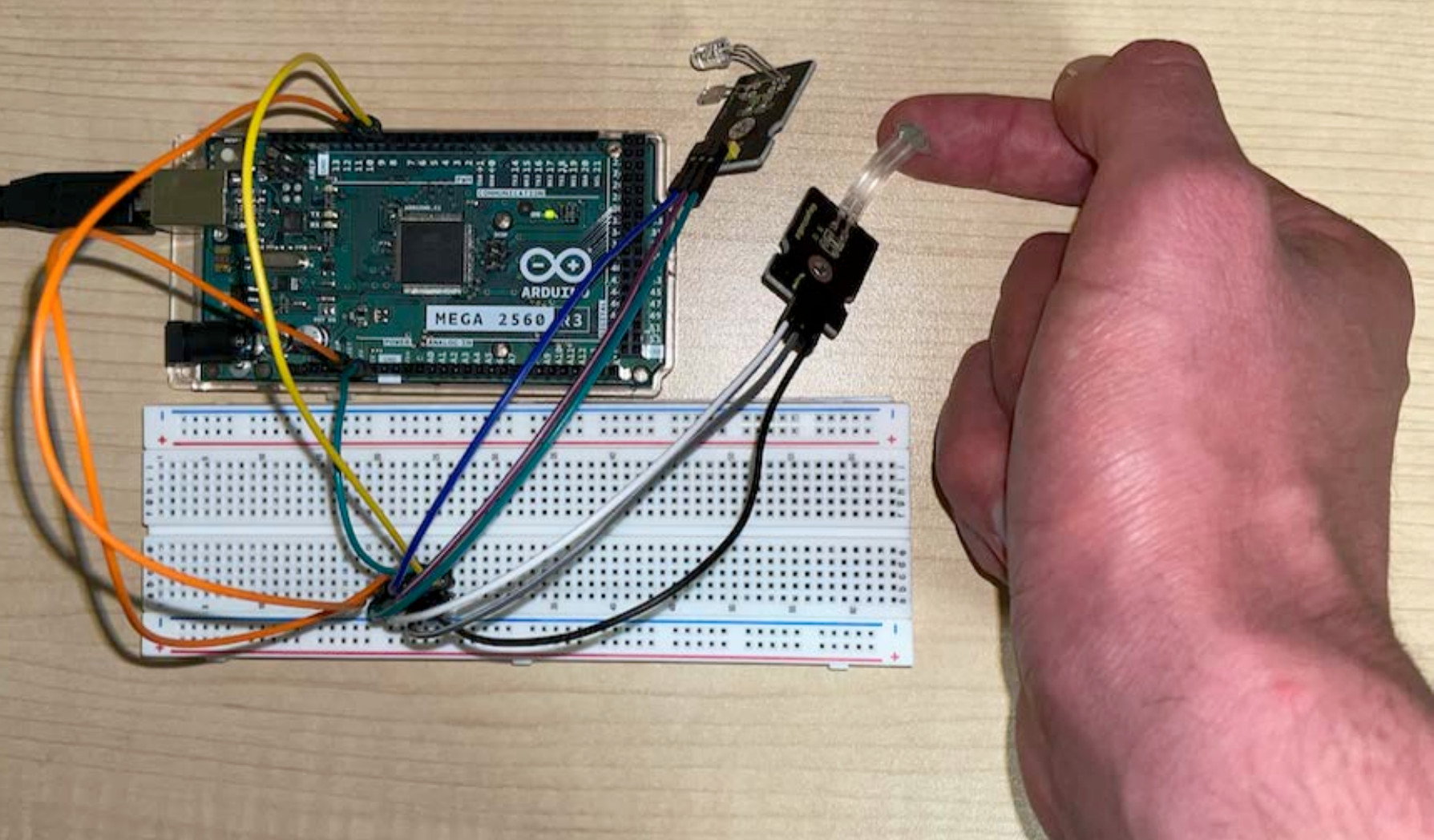}
		\label{fig:ss}
		\caption{Implementation Setup}
	\end{subfigure}
	
	\caption{\cco~Storage Impact on \comconstr~Delay and Energy Usage of \hases~schemes}
	\label{fig:energy-consumption}
\end{figure*}

\underline{\em Hardware and software configuration:}  
We implemented \hases~schemes on 8-bit MCU at the signer end using an Arduino Mega2560 board. This board features an 8-bit ATMega2560 MCU with 256KB flash memory, 8KB SRAM and 4KB EEPROM operating at 16MHz clock frequency.
We used the $\mu$NaCl open-source software library \footnote{\url{http://munacl.cryptojedi.org/index.shtml}}.
To demonstrate the cryptographic overhead in the low-end platforms, we use pressure sensor as our data source (see Fig. \ref{fig:energy-consumption}) to simulate sensory capabilities of twins equipped with sensors and actuators that replicate various bodily functions. We compared energy usage between one signature generation and a sensor sampling under different frequencies. Our goal is evaluating the trade-off between application and cryptographic overhead.

\underline{\em Performance Analysis:} Table \ref{tab:hases_8bit} showcases the performance comparison of \hases~schemes and their counterparts on the 8-bit MCU. We present the main takeaways as follows:

For a batch of $1024$ messages, \lahasessig~offer an aggregate signature generation that is  $9\times$ more efficient and only $6.5\times$ slower than an individual generation of an RSA and Ed25519 signature, respectively, computed on a single input. Unlike RSA, which introduces latency for batch processing, \lahases~is better suited for resource-constrained devices and massive data offload due its use of Curve25519, ensuring standard compliance and backward compatible.
Moreover, \lahases~offer a stateful digital signature that only trigger the pseudo-random number generator once during the key generation. Thus, it avoids the vulnerabilities that stems from the weak pseudo-random generators on resource-constrained devices.
Furthermore, \lahases~generate distinct commitments for each state, thereby ensuring tighter security reduction.

\begin{table*}[ht!]
	\caption{Performance analysis of \hases~schemes and counterparts on AVR ATmega2560 MCU}\label{tab:hases_8bit}
	\centering
	
	\resizebox{\textwidth}{!}{
		\begin{tabular}{|l || @{}c@{} | @{}c@{} | @{}c@{} | @{}c@{} | @{}c@{} | @{}c@{} | @{}c@{} | @{}c@{} | @{}c@{} | @{}c@{} | }
			\hline
			\textbf{Scheme} &  \specialcell[]{ \textbf{Signing} \\ \textbf{Time (sec)} }  &  \specialcell[]{ \textbf{Private} \\ \textbf{Key (KB)} }  & \specialcell[]{ \textbf{Signature}  \\ \textbf{Size (KB)} }  &  \specialcell[]{ \textbf{PQ} \\ \textbf{Promise} } & \specialcell[]{ \textbf{Standard} \\ \textbf{Compliant} } &  \specialcell[]{ \textbf{Backward} \\ \textbf{Compatible} }  &  \specialcell[]{ \textbf{Forward} \\ \textbf{Security} } & \specialcell[]{  \textbf{Agg} \\ \textbf{Capability} }  \\ \hline \hline

			Ed25519~\cite{Ed25519} & $1.42$ & $0.03$  & $0.06$  & $\times$ & $\checkmark$ & $\checkmark$ & $\times$ & $\times$  \\ \hline
			RSA-2048~\cite{Yavuz:TDSC:OutsourcedDB} & $83.26$  & $0.5$  & $0.25$  & $\times$ & $\checkmark$ & $\checkmark$ & $\times$ & $\checkmark$ \\ \hline 
			LA-HASES \hfill ($L\times$)  & $\boldsymbol{9.24}$ & $\boldsymbol{0.03}$  & $\boldsymbol{0.05}$  & $\times$ & $\checkmark$ & $\checkmark$ & $\times$ & $\checkmark$ \\ \hline \hline
			
			${}^\star$BLISS-\romannum{1}~\cite{espitau2017side} & $0.66$  & $2.00$ & $5.6$  & $\checkmark$ & $\times$ & $\times$ & $\times$ & $\times$ \\ \hline 
			
			PQ-HASES  & $\boldsymbol{0.01}$ & $\boldsymbol{0.03}$  & $\boldsymbol{0.5}$  & $\checkmark$ & $\checkmark$ & $\checkmark$ & $\checkmark$ & $\times$ \\ \hline \hline
			
			HY-HASES \hfill ($L\times$)  & $\boldsymbol{9.25}$ & $\boldsymbol{0.06}$  & $\boldsymbol{0.55}$  & $\checkmark$ & $\checkmark$ & $\checkmark$ & $\checkmark$ & $\checkmark$ \\ \hline
			
		\end{tabular}
	}
	
	\begin{tablenotes}[flushleft] \scriptsize{
			\item EPID \cite{boneh2019post} and SCB \cite{ouyang2021scb} are excluded as they are not inline with our system model (i.e., signers are low-end devices lacking secure enclaves). 
			\item $\star$ BLISS suffer from rejection sampling which results in devastating side-channel attacks \cite{espitau2017side}. $L$ denotes the batch size ($L=1024$).
			\item $\ddagger$ FI-BAF incur a large storage overhead at the verifier side that is linear w.r.t the number of items to be signed. 
		}
	\end{tablenotes}
\end{table*}

We identify only one PQ signature scheme, BLISS \cite{espitau2017side}, with benchmarking on 8-bit MCUs. Our \pqhases~is $66\times$ faster than BLISS-\romannum{1}, which is vulnerable from side-channel attacks due to rejection sampling \cite{espitau2017side}. \pqhases~is $142\times$ faster than standard Ed25519, which is neither PQ-secure nor FS. \pqhases~is standard compliant and backward-compatible. 

Among conventional-secure and PQ hybrids, \hyhases~stands as the sole feasible choice for DT-enabled platforms, especially on low-end devices, ensuring compact key sizes for continuous authentication.  \hyhases~is backward compatible and adheres to standards by leveraging SHA-256 and EC-based operations on Curve25519.  \hyhases~offers a highly efficient aggregate signature, complemented with a PQ and FS umbrella signature using minimal hash calls. Therefore, our experiments affirms \hyhases~as the ideal hybrid signature scheme for DT applications.

	\section{Conclusion} \label{sec:conclusion}
In this paper, we introduce Hardware-ASsisted Efficient Signatures (\hases) to address the critical requirements of lightweight digital signatures with exotic features in Digital Twins (DTs) applications. The \hases~schemes, including \pqhases, \lahases, and \hyhases, leverage a hardware-assisted cryptographic commitment construct oracle (CCO) to enable efficient verifiability without direct interaction with signers. These schemes offer features such as quantum-safe forward security, aggregate EC signatures, and hybrid conventional-secure and PQ combinations. By specifically targeting the resource limitations of low-end IoT devices commonly found in DT systems, \hases~ensure long-term security and breach resiliency. Experimental results demonstrate the superior signer efficiency of \pqhases~and \lahases~compared to their PQ and conventional-secure counterparts, respectively. 
The contributions of this work address the pressing need for lightweight digital signatures in DTs, providing both advanced security guarantees and efficiency. Through formal security proofs and comprehensive implementations, \hases~represent a significant advancement in ensuring the trustworthiness of sensitive information in DT applications. 
By open-sourcing the \hases~schemes, we invite further exploration, testing, and adaptation, facilitating progress in trustworthy DTs and enabling an effective transition into the PQ era with cryptographic agility. 

For potential future research directions, we aim to improve the forward security property in \pqhases. In fact, the forward security always incur an additional overhead that may be expensive, whether on the signer with additional computations and/or larger keys (e.g., \xmssmt \cite{cooper2020recommendation}) or at the verifier side with linear public keys (e.g., \pqhases). A potential research direction is to develop a forward security technique tailored for PQ digital signatures that provides efficiency for both the signer and the verifier. Additionally, we aim to further improve the security of \hases~by using a set of distributed servers to construct the one-time public keys. This technique eliminates the high risk of relying on a central root of trust. Furthermore, \hases~can be leveraged to generate the resource-intensive one-time commitments in multi-signatures, a process that traditionally involves multiple interactions during signing \cite{yavuz2023hardware}. We also foresee that \hases~can offer advantages to various applications (e.g., two-factor mobile authentication \cite{wang2016two}), when used as a signature primitive.

	\bibliographystyle{plain}
	\bibliography{AttilaYavuz, crypto-etc, agg-sig}

\begin{thebibliography}{10}

\bibitem{aloqaily2022integrating}
Moayad Aloqaily, Ouns Bouachir, Fakhri Karray, Ismaeel Al~Ridhawi, and
  Abdulmotaleb El~Saddik.
\newblock Integrating digital twin and advanced intelligent technologies to
  realize the metaverse.
\newblock {\em IEEE Consumer Electronics Mag.}, 2022.

\bibitem{anthoine2021dynamic}
Gaspard Anthoine, Jean-Guillaume Dumas, M{\'e}lanie de~Jonghe, Aude Maignan,
  Cl{\'e}ment Pernet, Michael Hanling, and Daniel~S Roche.
\newblock Dynamic proofs of retrievability with low server storage.
\newblock In {\em 30th USENIX Security Symposium (USENIX Security 21)}, pages
  537--554, 2021.

\bibitem{ateniese2008scalable}
Giuseppe Ateniese, Roberto Di~Pietro, Luigi~V Mancini, and Gene Tsudik.
\newblock Scalable and efficient provable data possession.
\newblock In {\em Proc of the 4th international conference on Security and
  privacy in communication netowrks}, pages 1--10, 2008.

\bibitem{bagchi2023post}
Prithwi Bagchi, Basudeb Bera, Ashok~Kumar Das, Sachin Shetty, Pandi
  Vijayakumar, and Marimuthu Karuppiah.
\newblock Post quantum lattice-based secure framework using aggregate signature
  for ambient intelligence assisted blockchain-based iot applications.
\newblock {\em IEEE Internet of Things Magazine}, 6(1):52--58, 2023.

\bibitem{barker2018recommendation}
Elaine Barker, Lily Chen, and Richard Davis.
\newblock Recommendation for key-derivation methods in key-establishment
  schemes.
\newblock {\em NIST Special Publication}, 800:56C, 2018.

\bibitem{behnia2021towards}
Rouzbeh Behnia and Attilla~Altay Yavuz.
\newblock Towards practical post-quantum signatures for resource-limited
  internet of things.
\newblock In {\em Proceedings of the 37th Annual Computer Security Applications
  Conference}, ACSAC '21, page 119–130, New York, NY, USA, 2021. Association
  for Computing Machinery.

\bibitem{Ed25519}
Daniel~J. Bernstein, Niels Duif, Tanja Lange, Peter Schwabe, and Bo-Yin Yang.
\newblock High-speed high-security signatures.
\newblock {\em Journal of Cryptographic Engineering}, 2(2):77--89, Sep 2012.

\bibitem{bernstein2019sphincs+}
Daniel~J Bernstein, Andreas H{\"u}lsing, Stefan K{\"o}lbl, Ruben Niederhagen,
  Joost Rijneveld, and Peter Schwabe.
\newblock The {SPHINCS+} signature framework.
\newblock In {\em Proc. of the ACM SIGSAC conf. on computer and comm.
  security}, pages 2129--2146, 2019.

\bibitem{bindel2017transitioning}
Nina Bindel, Udyani Herath, Matthew McKague, and Douglas Stebila.
\newblock Transitioning to a quantum-resistant public key infrastructure.
\newblock In {\em Post-Quantum Cryptography: 8th International Workshop,
  PQCrypto 2017, Utrecht, The Netherlands, June 26-28, 2017, Proceedings 8},
  pages 384--405. Springer, 2017.

\bibitem{boldyreva2007ordered}
Alexandra Boldyreva, Craig Gentry, Adam O'Neill, and Dae~Hyun Yum.
\newblock Ordered multisignatures and identity-based sequential aggregate
  signatures, with applications to secure routing.
\newblock In {\em Proceedings of the 14th ACM conference on Computer and
  communications security}, pages 276--285, 2007.

\bibitem{boneh2019post}
Dan Boneh, Saba Eskandarian, and Ben Fisch.
\newblock Post-quantum epid signatures from symmetric primitives.
\newblock In {\em Topics in Cryptology--CT-RSA 2019: The Cryptographers' Track
  at the RSA Conference 2019, San Francisco, CA, USA, March 4--8, 2019,
  Proceedings}, pages 251--271. Springer, 2019.

\bibitem{boneh2020one}
Dan Boneh and Sam Kim.
\newblock One-time and interactive aggregate signatures from lattices.
\newblock {\em preprint}, 2020.

\bibitem{BLS:2004:Boneh:JournalofCrypto}
Dan Boneh, Ben Lynn, and Hovav Shacham.
\newblock Short signatures from the weil pairing.
\newblock {\em J. Cryptol.}, 17(4), 2004.

\bibitem{boudgoust2023sequential}
Katharina Boudgoust and Akira Takahashi.
\newblock Sequential half-aggregation of lattice-based signatures.
\newblock {\em Cryptology ePrint Archive}, 2023.

\bibitem{chamola2021information}
Vinay Chamola, Alireza Jolfaei, Vaibhav Chanana, Prakhar Parashari, and Vikas
  Hassija.
\newblock Information security in the post quantum era for 5g and beyond
  networks: Threats to existing cryptography, and post-quantum cryptography.
\newblock {\em Computer Communications}, 176:99--118, 2021.

\bibitem{chen2022lfs}
Xue Chen, Shiyuan Xu, Yunhua He, Yu~Cui, Jiahuan He, and Shang Gao.
\newblock Lfs-as: lightweight forward secure aggregate signature for e-health
  scenarios.
\newblock In {\em ICC 2022-IEEE Intern. Conf. on Communications}, pages
  1239--1244. IEEE, 2022.

\bibitem{chen2022half}
Yanbo Chen and Yunlei Zhao.
\newblock Half-aggregation of schnorr signatures with tight reductions.
\newblock In {\em European Symposium on Research in Computer Security}, pages
  385--404. Springer, 2022.

\bibitem{cooper2020recommendation}
David~A Cooper, Daniel~C Apon, Quynh~H Dang, Michael~S Davidson, Morris~J
  Dworkin, Carl~A Miller, et~al.
\newblock Recommendation for stateful hash-based signature schemes.
\newblock {\em NIST Special Publication}, 800:208, 2020.

\bibitem{costan2016sanctum}
Victor Costan, Ilia Lebedev, and Srinivas Devadas.
\newblock Sanctum: Minimal hardware extensions for strong software isolation.
\newblock In {\em 25th USENIX Security Symposium (USENIX Security 16)}, pages
  857--874, 2016.

\bibitem{SchnorrQ}
Craig Costello and Patrick Longa.
\newblock Schnorrq: Schnorr signatures on fourq.
\newblock Technical report, MSR Tech Report, 2016. Available at: https://www.
  microsoft. com/en-us/research/wp-content/uploads/2016/07/SchnorrQ. pdf, 2016.

\bibitem{crockett2019prototyping}
Eric Crockett, Christian Paquin, and Douglas Stebila.
\newblock Prototyping post-quantum and hybrid key exchange and authentication
  in tls and ssh.
\newblock {\em Cryptology ePrint Archive}, 2019.

\bibitem{dobraunig2021ascon}
Christoph Dobraunig, Maria Eichlseder, Florian Mendel, and Martin
  Schl{\"a}ffer.
\newblock Ascon v1. 2: Lightweight authenticated encryption and hashing.
\newblock {\em Journal of Cryptology}, 34:1--42, 2021.

\bibitem{drijvers2020pixel}
Manu Drijvers, Sergey Gorbunov, Gregory Neven, and Hoeteck Wee.
\newblock Pixel: Multi-signatures for consensus.
\newblock In {\em USENIX Security Symposium}, pages 2093--2110, 2020.

\bibitem{ducas2018crystals}
L{\'e}o Ducas, Eike Kiltz, Tancrede Lepoint, Vadim Lyubashevsky, Peter Schwabe,
  Gregor Seiler, and Damien Stehl{\'e}.
\newblock Crystals-dilithium: A lattice-based digital signature scheme.
\newblock {\em IACR Transactions on Cryptographic Hardware and Embedded
  Systems}, pages 238--268, 2018.

\bibitem{el2022benchmarking}
Muhammad El-Hindi, Tobias Ziegler, Matthias Heinrich, Adrian Lutsch, Zheguang
  Zhao, and Carsten Binnig.
\newblock Benchmarking the second generation of intel sgx hardware.
\newblock In {\em Data Management on New Hardware}, pages 1--8. 2022.

\bibitem{el2021potential}
Abdulmotaleb El~Saddik, Fedwa Laamarti, and Mohammad Alja'Afreh.
\newblock The potential of digital twins.
\newblock {\em IEEE Instrumentation \& Measurement Magazine}, 24(3):36--41,
  2021.

\bibitem{espitau2017side}
Thomas Espitau, Pierre-Alain Fouque, Beno{\^\i}t G{\'e}rard, and Mehdi
  Tibouchi.
\newblock Side-channel attacks on bliss lattice-based signatures: Exploiting
  branch tracing against strongswan and electromagnetic emanations in
  microcontrollers.
\newblock In {\em Proceedings of the 2017 ACM SIGSAC Conference on Computer and
  Communications Security}, pages 1857--1874, 2017.

\bibitem{fouque2018falcon}
Pierre-Alain Fouque, Jeffrey Hoffstein, Paul Kirchner, Vadim Lyubashevsky,
  Thomas Pornin, Thomas Prest, Thomas Ricosset, Gregor Seiler, William Whyte,
  Zhenfei Zhang, et~al.
\newblock Falcon: Fast-fourier lattice-based compact signatures over ntru.
\newblock {\em Submission to the NIST’s post-quantum cryptography
  standardization process}, 36(5), 2018.

\bibitem{XMSS}
Andreas Huelsing, Denis Butin, Stefan-Lukas Gazdag, Joost Rijneveld, and Aziz
  Mohaisen.
\newblock {XMSS: eXtended Merkle Signature Scheme}.
\newblock RFC 8391, May 2018.

\bibitem{joseph2022transitioning}
David Joseph, Rafael Misoczki, Marc Manzano, Joe Tricot, Fernando~Dominguez
  Pinuaga, Olivier Lacombe, Stefan Leichenauer, Jack Hidary, Phil Venables, and
  Royal Hansen.
\newblock Transitioning organizations to post-quantum cryptography.
\newblock {\em Nature}, 605(7909):237--243, 2022.

\bibitem{katz2020introduction}
Jonathan Katz and Yehuda Lindell.
\newblock {\em Introduction to modern cryptography}.
\newblock CRC press, 2020.

\bibitem{lang2021informer}
Fan Lang, Wei Wang, Lingjia Meng, Qiongxiao Wang, Jingqiang Lin, and Li~Song.
\newblock Informer: Protecting intel sgx from cross-core side channel threats.
\newblock In {\em Intern. Conf. on Information and Communications Security},
  pages 310--328, 2021.

\bibitem{li2020permissioned}
Tian Li, Huaqun Wang, Debiao He, and Jia Yu.
\newblock Permissioned blockchain-based anonymous and traceable aggregate
  signature scheme for industrial internet of things.
\newblock {\em IEEE Internet of Things Journal}, 8(10):8387--8398, 2020.

\bibitem{ForwardSecure_MMM_02}
T.~Malkin, D.~Micciancio, and S.~K. Miner.
\newblock Efficient generic forward-secure signatures with an unbounded number
  of time periods.
\newblock In {\em Proc. of the 21th International Conference on the Theory and
  Applications of Cryptographic Techniques ({EUROCRYPT} '02)}, pages 400--417.
  Springer-Verlag, 2002.

\bibitem{mao2016resource}
Hongzi Mao, Mohammad Alizadeh, Ishai Menache, and Srikanth Kandula.
\newblock Resource management with deep reinforcement learning.
\newblock In {\em Proceedings of the 15th ACM workshop on hot topics in
  networks}, pages 50--56, 2016.

\bibitem{nist-4th-round}
NIST.
\newblock {PQC Standardization Process: Announcing Four Candidates to be
  Standardized, Plus Fourth Round Candidates}.
\newblock
  \url{https://csrc.nist.gov/News/2022/pqc-candidates-to-be-standardized-and-round-4}.
\newblock Accessed: July 14, 2022.

\bibitem{Yavuz:HASES:ICC2023}
Saif~E. Nouma, , and Attila~A. Yavuz.
\newblock Post-quantum forward-secure signatures with hardware-support for
  internet of things.
\newblock IEEE International Conference on Communications (ICC), page 1–7.
  IEEE, 2023.

\bibitem{nouma2023practical}
Saif~E Nouma and Attila~A Yavuz.
\newblock Practical cryptographic forensic tools for lightweight internet of
  things and cold storage systems.
\newblock In {\em Proc. of the 8th ACM/IEEE Conf. on Internet of Things Design
  and Implementation}, pages 340--353, 2023.

\bibitem{ott2019identifying}
David Ott, Christopher Peikert, et~al.
\newblock Identifying research challenges in post quantum cryptography
  migration and cryptographic agility.
\newblock {\em arXiv preprint arXiv:1909.07353}, 2019.

\bibitem{ouyang2021scb}
Wenyi Ouyang, Qiongxiao Wang, Wei Wang, Jingqiang Lin, and Yaxi He.
\newblock {SCB: Flexible and Efficient Asymmetric Computations Utilizing
  Symmetric Cryptosystems Implemented with Intel SGX}.
\newblock In {\em 2021 IEEE International Performance, Computing, and
  Communications Conference (IPCCC)}, pages 1--8, 2021.

\bibitem{Yavuz:CNS:2019}
Muslum~Ozgur Ozmen, Rouzbeh Behnia, and Attila~A. Yavuz.
\newblock Energy-aware digital signatures for embedded medical devices.
\newblock In {\em 7th {IEEE} Conf. on Communications and Network Security
  ({CNS}), Washington, D.C., USA, June}, 2019.

\bibitem{paul2020towards}
Sebastian Paul and Patrik Scheible.
\newblock Towards post-quantum security for cyber-physical systems: Integrating
  pqc into industrial m2m communication.
\newblock In {\em Computer Security--ESORICS 2020: 25th European Symposium on
  Research in Computer Security, ESORICS 2020, Guildford, UK, September 14--18,
  2020, Proceedings, Part II 25}, pages 295--316. Springer, 2020.

\bibitem{SecProofSigScheme96Euro}
D.~Pointcheval and J.~Stern.
\newblock Security proofs for signature schemes.
\newblock In {\em Proc. of the 15th International Conference on the Theory and
  Application of Cryptographic Techniques ({EUROCRYPT} '96)}, pages 387--398.
  Springer-Verlag, 1996.

\bibitem{qassim2017post}
Yousef Qassim, Mario~Edgardo Maga{\~n}a, and Attila Yavuz.
\newblock Post-quantum hybrid security mechanism for mimo systems.
\newblock In {\em Computing, Networking and Communications (ICNC), 2017
  International Conference on}, pages 684--689. IEEE, 2017.

\bibitem{Reyzin2002}
Leonid Reyzin and Natan Reyzin.
\newblock Better than {BiBa}: Short one-time signatures with fast signing and
  verifying.
\newblock In {\em Information Security and Privacy: 7th Australasian
  Conference}, pages 144--153, July 2002.

\bibitem{Yavuz:CORE:CNS:2020}
Efe U.~A. Seyitoglu, Attila~A. Yavuz, and Muslum~O. Ozmen.
\newblock Compact and resilient cryptographic tools for digital forensics.
\newblock In {\em 2020 IEEE Conference on Communications and Network Security
  (CNS)}, pages 1--9, 2020.

\bibitem{shaw2022post}
Surbhi Shaw and Ratna Dutta.
\newblock Post-quantum secure identity-based signature achieving forward
  secrecy.
\newblock {\em Journal of Information Security and Applications}, 69:103275,
  2022.

\bibitem{shengli2021human}
Wei Shengli.
\newblock Is human digital twin possible?
\newblock {\em Computer Methods and Programs in Biomedicine Update}, 1:100014,
  2021.

\bibitem{Shor-algo}
Peter~W. Shor.
\newblock Polynomial-time algorithms for prime factorization and discrete
  logarithms on a quantum computer.
\newblock {\em SIAM Review}, 41(2):303--332, 1999.

\bibitem{silva2022power}
Rangana~De Silva, Iranga Navaratna, Malitha Kumarasiri, Janaka Alawatugoda, and
  Chuah~Chai Wen.
\newblock On power analysis attacks against hardware stream ciphers.
\newblock {\em Intern. J. of Information and Computer Security},
  17(1-2):21--35, 2022.

\bibitem{vallent2021efficient}
Thokozani~F. Vallent, Damien Hanyurwimfura, and Chomora Mikeka.
\newblock Efficient certificateless aggregate signature scheme with conditional
  privacy-preservation for vehicular adhoc networks enhanced smart grid system.
\newblock {\em Sensors}, 2021.

\bibitem{wang2011privacy}
Cong Wang, Sherman~SM Chow, Qian Wang, Kui Ren, and Wenjing Lou.
\newblock Privacy-preserving public auditing for secure cloud storage.
\newblock {\em IEEE transactions on computers}, 62(2):362--375, 2011.

\bibitem{wang2014anonymous}
Ding Wang, Debiao He, Ping Wang, and Chao-Hsien Chu.
\newblock Anonymous two-factor authentication in distributed systems: Certain
  goals are beyond attainment.
\newblock {\em IEEE TDSC}, 12(4):428--442, 2014.

\bibitem{wang2016two}
Ding Wang and Ping Wang.
\newblock Two birds with one stone: Two-factor authentication with security
  beyond conventional bound.
\newblock {\em IEEE transactions on dependable and secure computing},
  15(4):708--722, 2016.

\bibitem{wang2021quantum2fa}
Qingxuan Wang, Ding Wang, Chi Cheng, and Debiao He.
\newblock Quantum2fa: efficient quantum-resistant two-factor authentication
  scheme for mobile devices.
\newblock {\em IEEE Transactions on Dependable and Secure Computing}, 2021.

\bibitem{Yavuz:TDSC:OutsourcedDB}
Attila~A. Yavuz.
\newblock Immutable authentication and integrity schemes for outsourced
  databases.
\newblock {\em {IEEE} Trans. Dependable Sec. Comput.}, 15(1):69--82, 2018.

\bibitem{yavuz2022frog}
Attila~A Yavuz and Rouzbeh Behnia.
\newblock Frog: Forward-secure post-quantum signature.
\newblock {\em arXiv preprint arXiv:2205.07112}, 2022.

\bibitem{Yavuz:2012:TISSEC:FIBAF}
Attila~A. Yavuz, Peng Ning, and Michael~K. Reiter.
\newblock {BAF} and {FI-BAF}: Efficient and publicly verifiable cryptographic
  schemes for secure logging in resource-constrained systems.
\newblock {\em ACM Trans on Information System Security}, 15(2), 2012.

\bibitem{yavuz2022distributed}
Attila~A Yavuz, Saif~E Nouma, Thang Hoang, Duncan Earl, and Scott Packard.
\newblock Distributed cyber-infrastructures and artificial intelligence in
  hybrid post-quantum era.
\newblock In {\em 2022 IEEE 4th International Conference on Trust, Privacy and
  Security in Intelligent Systems, and Applications (TPS-ISA)}, pages 29--38.
  IEEE, 2022.

\bibitem{yavuz2023hardware}
Attila~Altay Yavuz and Saif Nouma.
\newblock Hardware supported authentication and signatures for wireless,
  distributed and blockchain systems, September~28 2023.
\newblock US Patent App. 18/188,749.

\bibitem{zhu2009smart}
Haojin Zhu, Xiaodong Lin, Rongxing Lu, Yanfei Fan, and Xuemin Shen.
\newblock Smart: A secure multilayer credit-based incentive scheme for
  delay-tolerant networks.
\newblock {\em IEEE transactions on vehicular technology}, 58(8):4628--4639,
  2009.

\end{thebibliography}

\section*{APPENDIX A}
\label{appendixA}

Our schemes rely on HORS~\cite{Reyzin2002}  and Schnorr \cite{SchnorrQ} digital signatures that are defined as below. 

\begin{definition} \label{alg:hors}
	The Hash to Obtain Random Subset ($\hors$) signature~\cite{Reyzin2002} is a tuple of three algorithms $(\kg,\ssig,\ver)$ defined as follows:
	\begin{enumerate}[\indent -]
		\item $ \underline{ (\sk, \pk, I) \as \horskg(1^{\kappa}) } $: Given the security parameter $\kappa$, it first generates  $ I \as (l,k,t) $, and then  $t$ random $l$-bit strings $\{s_i\}_{i=1}^t$ and  $\{ v_i \as f(s_i) \}_{i=1}^{t}$. It sets $ \sk \as \{ s_i \}_{i = 1}^t$ and $\pk \as \{ v_i \}_{i = 1}^t$.
		\item $\underline{\sigma\as \horssig(\sk, M)}$: Given $(\sk,M)$, it computes $h \as H_0(M)$, splits it as $\{h_j\}_{j=1}^k$ where $|h_j| = \log{t}$ and interprets them as integers $\{i_j\}_{j=1}^k$. It sets $\sigma \as   \{ s_{i_j} \}_{j = 1}^k $.
		\item $ \underline{ b \as \horsver( \pk, M, \sigma) } $: Given \pk, $M$,  and $\sigma$, it computes $\{i_j\}_{j=1}^k$ as in  $\horssig(.)$~and checks if $ f(s'_j) = v_{i_j}$, $j=1,\ldots,k$, returns $b=1$, else  $b=0$.
	\end{enumerate}
\end{definition}

\begin{definition} \label{alg:Schnorr}
	The \sch~signature ~\cite{SchnorrQ} is a tuple of three algorithms defined as follows:
	\begin{enumerate}[\indent -]
		\item $ \underline{ (y, Y, I) \as \schkg(1^{\kappa}) } $: Given the security parameter $\kappa$, it first generates large primes $q$ and $p>q$ such that $q|(p-1)$. Select a generator $\alpha$ of the subgroup $G$ of order $q$ in $\mathbb{Z}_{p}^{*}$.
		\item $\underline{\sigma\as \schsig(y, M)}$: Given $(y, M)$, it generates private/public commitment $(r \Rq, R \as \alpha^r \mod p)$. It computes an ephemeral key $e \as H_0(M \| R)$ and the signature $s\as r-e \cdot y$. It sets $\sigma \as \langle s, e \rangle$. 
		\item $ \underline{ b \as \schver( Y, M, \sigma) }$:  Given $Y$, $M$, and $\sigma$, it computes $R' \as Y^e \cdot \alpha^s \mod p$. It checks if $e = H(M \| R) $, then it returns $b=1$, otherwise $b=0$.
	\end{enumerate}
\end{definition}

The Discrete Logarithm Problem (DLP) is defined as below:

\begin{definition}
	\label{def.dlp}
	Let $\mathbb{G}$ be a cyclic group of order $q$, let $\alpha$ be a generator of $\mathbb{G}$, and let \dlp~attacker \A~be an algorithm that returns an integer in $\mathbb{Z}_q$.  
	We consider the following experiment:
	
	Experiment $Expt_{\mathbb{G}, \alpha}^{DL}(\mathcal{A})$:
	\newline 
	\tab $b \Ra \mathbb{Z}_q^*$, $B \as \alpha^{b} \mod q$,
	$b' \as \mathcal{A}(B)$,
	\newline 
	\tab If $\alpha^{b'} \mod p = B$ then return 1, else return 0
	\newline
	The DL-advantage of \A~in this experiment is defined as:
	~$\advAdl = Pr[ Expt_{\mathbb{G}, \alpha}^{DL}(\mathcal{A}) = 1]$
	\newline
	The \dl~advantage of $(\mathbb{G}, \alpha)$ in this experiment is defined as follows:
	~$\advdl = \underset{\mathcal{A}}{\max}\{\advAdl\}$, where the maximum is over all \A~having time complexity $t$.
\end{definition}

We implemented our schemes in the EC domain due to the small key sizes. Specifically, we used the curve of Ed25519 \cite{Ed25519} that offers efficient arithmetics. The security of Ed25519 relies on Elliptic Curve Discrete Logarithm Problem (ECDLP), which is the EC variant of DLP in Definition \ref{def.dlp}.

\section*{APPENDIX B}
\label{appendixB}

We provide security proof of \pqhases~scheme as below.

\begin{theoremp}{5.1} \label{theorem1}
	$\advpqhases  \le  J \cdot \advhors$, where $\mathcal{O}(t')=\mathcal{O}(t) + J  \cdot H$.
\end{theoremp}

\noindent {\em Proof:}  \F~is given the challenge public key $PK'$,  where $(sk',PK',I) \leftarrow \horskg(1^{\kappa})$ . 

\textit{\vspace{2mm} \noindent \underline{{\em Algorithm $\mathcal{F}^{\ro(\cdot), \hors_{sk'}(\cdot),\Breakin}(\pk')$}}}: \F~is run per Def. \ref{def:FEUCMA:HMU} as follows.

$\bullet$~\underline{{\em Setup:}} 
\F~maintains \lm, and \ls~to track the query results in the experiment duration.
\lm~is a list of queried messages to the signing oracle $\pqhasessig_{\sk'}$. 
\ls~is a signature list to record answers given by $\pqhasessig_{\sk'}$. 
\lh~is a hash list in form of pairs $\{(M_l, k) : h_l \}$, where $M_l$ is the $l^{th}$ data item queried to $\ro$ and $h_l$ is its associated $\ro$ answer. $k=0$ denotes the selection of hash function $H$ or $k=1$ in case of \prf~function.
\A~selects an $ID_n$ and gives it to \F. If $ID_n \notin \vec{ID}$, then \F~{\em aborts}, else continues. The index $n$, for the user $ID_n$, is omitted for brevity.  \F~uses a function \hsim~to model $H$ as a random oracle \ro: If $\exists (M, k) : \lh[M,k]=h $ then \hsim~returns $h$. Else, it returns $h \as \{0,1\}^{\kappa}$ as the answer for $H_k$, insert a new pair $\lh(M, k) \as h$ and update $l \as l+1$.   \F~selects a target forgery index $w  \in [1,J]$ and sets $C_w \as \pk'$.  \F~generates $\msk \Ra \{0,1\}^\kappa$,  and computes  $\{\sk_j\}_{j=1, j\neq w }^J$ and  $\{C_j,\sigma_{C_j}\}_{j=1, j\neq w}^J$ as in \pqhaseskg~and \pqhasescomconstr~via \msk~and \csk, where $H$ and \prf~calls are simulated with \hsim.

$\bullet$~\underline{{\em Queries:}} \F~handles \A's queries as follows.

{\em (1) \ro}: \A~queries \ro~on $q_s'$ messages of her choice. When \A~queries $M$ for a hash function $H_{k=0,1}$, \F~returns $h \as \hsim(M, k, l, \lh)$.  

{\em (2) \Sigj}: If \A~queries \F~on $M_w$ then \F~returns $\sigma_j \as \hors_{sk'}(M_j)$  by querying $\hors_{\sk'}(.)$~signing oracle. Otherwise, \F~returns $\sigma_j \as \pqhasessig(\sk_j \in \lh ,M_j)$ ($\sk_j$ is a \prf~output, and therefore it is available in \lh). 

{\em (3) $\comconstr_{msk}(.)$}: If \A~queries \F~on $\text{w}^{\text{th}}$ commitment, then \F~returns $C_w=\pk'$. Else, \F~returns $C_j$ from \lh~that is computed via \pqhasescomconstr~algorithm under \msk. 

{\em (4) \Breakin}: If $j = J$ then \F~rejects the query (all private keys were used) and proceed to the forgery phase. If $1<j \leq w$ then \F~{\em aborts}, else it returns $\sk_j \in \lh$ and ends the experiment. 

$\bullet$~\underline{{\em Forgery:}} \A~outputs a forgery $(M^{*},\sigma^{*})$ on $PK^{*}$. \F~wins the experiments if \A~wins the experiments by producing a valid forgery on $PK_w$. That is, \F~returns 1 if $PK^{*} = PK_w$,  $M^{*}$ was not queried to \Sigj, and $1=\horsver(PK^{*}, M^{*},\sigma^{*})$. Otherwise, \F~returns 0 and {\em aborts}.

$\bullet$~\underline{{\em Success Probability and Indistinguishability Argument}}: \sloppy Assume that \A~wins \FEUCMA~experiment against \pqhases~with the probability \advpqhases. \F~wins the \EUCMA~experiments against \hors~if and only if \A~produces a forgery on the challenge public key $PK_w$ and does not abort during the experiment. Since  $w \in [1,J]$ is randomly selected, the success probability of \A~can be bounded to that of \F~as:~~~  $\advpqhases  \le  J \cdot \advhors$
\vspace{2pt}

\F's transcripts in all lists are identical to the real execution except that $PK_w$ is replaced with \hors~public key $PK'$. Hence, $\sk_w = sk'$ is not part of hash chain generated from $\sk_1 $  or \msk. As in \hors,  $H$ is a random oracle. Thus, $w^\text{th}$ element of lists in  $\mathcal{A}_{\mathcal{R}}$  and $\mathcal{A}_{\mathcal{S}}$ have identical distributions. \hfill $\blacksquare$


\vspace{3mm}
We give the security proof of \lahases~as below.

\vspace{1mm}
\begin{theoremp}{5.2} \label{theorem2}
	$\advlahases  \le  \advdll$, where $t' = \mathcal{O}(t) + \mathcal{O}(\kappa^3 + L \cdot \RNG)$.
\end{theoremp}

\noindent {\em Proof:} Let \A~be a \lahases~attacker. We construct a {\em DL-attacker} \F~that uses $\mathcal{A}$ as a sub-routine. That is, we set $(b\Rq,~B\as \alpha^{b} \bmod p)$ as in Def.~\ref{def.dlp} and then run the simulator \F~by Def.~\ref{Def:AEUCMA:HAMU} (i.e., \AEUCMA) as follows:

\vspace{2mm} \noindent \underline{{\em Algorithm $\mathcal{F}(B)$}}: \F~is run per Definition \ref{Def:AEUCMA:HAMU}.
\begin{enumerate}[ ]
	\setlength{\itemsep}{2pt}
	\setlength{\parskip}{0pt}
	\setlength{\parsep}{0pt}
	
	\item \underline{{\em Setup:}} 
	\F~maintains \lh, \lm, \ls, \lc~and \lr~to keep track of query results in the duration of the experiments. 
	\lh~is a hash list in form of pairs $\{(M_l, k) : h_l \}$, where $(M_l,h_l)$ represents the $l^{th}$ data item queried to $\ro$ and its corresponding $\ro$ answer, respectively. $k=0$ and $k=1$ denotes the selection of hash function $H$ and \prf, respectively.
	\lm~is a list containing vectors of messages, each of its elements $\lm[j]$ is a message vector $\overrightarrow{M}_j=\{m_{j}^{\ell}\}_{\ell=1}^L$ (i.e., the $j^{\text{th}}$ batch query to $\lahasessig_{sk}$).
	\ls~is a signature list, used to record answers given by $\lahasessig_{\sk}$.
	\lr~is a list, used to record the randomly generated variables. 
	$\lr[0,0,0]$ and $\lr[1,j,\ell]$ refer to the simulated private key $z$ (generated by \ro) and  $s_{j}^\ell$ during the message element index $\ell$ of the batch query $j$, respectively.

	\vspace*{2mm}
	\begin{itemize}
		\item \F~sets the public key and the system-wide parameters as follows: 
		$Y \as B, z \Rq$ and add to \lr~as $\lr[0,0,0] \as z$. Set $I \as (p,q,\alpha, J, L)$, and  initialize the counters $(l \as 0,j \as 1)$. 
		\item \A~selects an $ID$ and gives it to $\mathcal{F}$. If $ID \notin \vec{ID}$, \F~aborts, else continues (user index $n$ is omitted).
		\item $h \as \hsim(M, k, l, \lh)$: \F~uses a function \hsim~that works as a random oracle, \ro. Note that the hash and \prf~functions $H_{k=0,1}$ are modeled as random oracles. If $\exists (M, k) : \lh[M,k]=h $ then \hsim~returns $h$. Otherwise, return $h \Rq$ as the answer for $H_k$, insert a new pair $\lh(M, k) \as h$ and update $l \as l+1$. 
	\end{itemize}

	\item \underline{Execute $\mathcal{A}^{\ro, \lahasessig_{sk}(\cdot), \lahasescomconstr(\cdot)}(\pk)$}: 
	
	$\bullet$~{\em Queries:} \F~handles \A's queries as follows:
	\begin{enumerate}[(1)]
		\item \ro: \A~queries \ro~on $q_s'$ messages of her choice. When \A~queries $M$ for a hash function $H_{k=0,1}$, \F~returns $h \as \hsim(M, k, l, \lh)$.
		
		\item \Sig: \A~queries the $\lahasessig(.)$~oracle on $J$ data batches of her choice $\vec{M_j}=\{m_{j}^\ell \}_{\ell=1}^L, \forall j \in [1,J]$. If $j>J$, \F~rejects the query (exceed limit), else it continues as follows:
		\begin{enumerate}[a)]
			\item Compute $z \as \lr[0,0,0]$~, $x_{j} \as \hsim(z \| j, 0, l, \lh)$,  and set $s_j^{1,0} \as 0$
			\item \textbf{for} $\ell=1, \ldots, L$ \textbf{do}
			\begin{enumerate}[i)]
				\item Compute $x_j^{\ell} \as \hsim(x_j \| \ell, 0, l, \lh)$ and $e_j^\ell \as \hsim(x_j^\ell, 2, l, \lh)$
				\item If $(m_j^{\ell} \| x_j^{\ell}, 2) \in \lh$ then \F~aborts. Otherwise, it computes $\hsim(m_j^{\ell} \| x_j^{\ell}, 2, l, \lh)$
				\item If $\exists (j,\ell): \lr[1,j,\ell]=s_j^\ell$ then $s_j^\ell \as \lr[2,j,\ell]$. Otherwise, $s_j^\ell \Rq$ and insert it into \lr~as $\lr[1,j,\ell] \as s_j^\ell $. Compute $s_j^{1,\ell} \as s_j^{1, \ell-1} + s_j^{\ell} \mod q$
				
			\end{enumerate}			
			\item \F~sets $\sigma_j^{1,L} \as \langle s_j^{1,L}, x_j , ID, j \rangle$, insert $(\vec{M}_j, \sigma_j^{1,L})$ to (\lm, \ls) and increment $j\as j+1$
		\end{enumerate}
		
		\item \comconstrmsk: \A~queries \lahasescomconstr~oracle on user  $ID$ and counter $j$ of her choice. If $j \notin [1,J]$ or $ID \notin \vec{ID}$, \F~reject the query, else it continues as follows:
		\begin{enumerate}[a)]
			\item $e_j^{1,0} \as 0$ and $s_j^{1,0} \as 0$. Retrieve $z \as \lr[0,0,0]$ and Compute $x_j \as \hsim[z \| j, 0, l, \lh]$
			\item \textbf{for} $\ell=1, \ldots, L$ \textbf{do}
			\begin{enumerate}[i)]
				\item Compute $x_j^\ell \as \hsim(x_j \| \ell, 0, l, \lh)$ and $e_j^\ell \as \hsim(x_j^\ell, 2, l, \lh)$
				\item If $\exists (j,\ell): \lr[2,j,\ell]=s_j^\ell$ then $s_j^\ell \as \lr[1,j,\ell]$. Otherwise, generate $s_j^\ell \Rq$ and insert $\lr[1,j,\ell] \as s_j^\ell $
			\end{enumerate}
			\item Set $R_j^{1,L} \as Y^{ \sum_{\ell=1}^{L} e_j^\ell }  \cdot \alpha^{ \sum_{\ell=1}^{L} s_j^\ell } \mod p$, return and insert $C_j^{1,L} \as R_j^{1,L}$ to \lc~as $\lc[j] \as C_j^{1,L}$.
		\end{enumerate}
	\end{enumerate}

	\begin{list}{-}{}
		\vspace{2mm}
		\item   \sloppy \underline{Forgery of \A:} Eventually, \A~outputs forgery on \pk~as $(\vec{M}_{j}^*,$ ${\sigma^*}_{j}^{1,L})$, for $j\in [1,J]$, where $\vec{M}_{j}^*= \{ {m^*}_{j}^{1}, \ldots, {m^*}_{j}^{L} \} $ and ${\sigma^*}_{j}^{1,L}=\langle {s^*}_{j}^{1,L}, {x}_{j}^*, ID \rangle$. By definition \ref{Def:AEUCMA:HAMU}, \A~wins \AEUCMA~experiment for \lahases~if $\lahasesver(\langle \pk,{C}_{j}^{1,L} \rangle ,\vec{M}_{j}^*,{\sigma^*}_{j}^{1,L})=1$ and  $\vec{M}_{j}^* \notin \lm$ hold, where ${C}_{j}^{1,L} \as \lahasescomconstr_{\msk}(ID,j)$. If these conditions hold, \A~returns $1$, otherwise $0$.
		
		\vspace{2mm}
		\item \underline{Forgery of \F}: If \A~loses in the \AEUCMA~experiment for \lahases, \F~also loses in the \dl~experiment, and thus \F~{\em aborts} and returns $0$. Else, if $\vec{M}_{j}^* \in \lm$ then \F~{\em aborts} and returns $0$ (i.e., \A~wins the experiment without querying \ro~oracle).  Otherwise, \F~continues as follows:
		
		$R_{j}^{1,L} \equiv Y^{e_{j}^{1,L}} \cdot \alpha^{s_{j}^{1,L}} \mod p$ holds for the aggregated variables $(R_{j}^{1,L},e_{j}^{1,L},s_{j}^{1,L})$. That is, \F~retrieves $z \as \lr[0,0,0]$ and computes $x_j \as \hsim(z\|j,2,l,\lh)$. Then, it computes $e_j^{1,L} \as \sum_{\ell=1}^L \hsim(x_j^\ell, 2,l,\lh)$ where $x_j^\ell=\hsim(x_j \| \ell,0,l,\lh), \forall \ell \in [1,L]$. Also, \F~computes $s_j^{1,L} \as \sum_{\ell=1}^L \lr[1,j,\ell]$.
		
		Moreover, $\lahasesver(\vec{M}_{j}^* , {\sigma^*}_{j}^{1,L} )=1$ holds, and therefore $R_{j}^{1,L} \equiv Y^{{e^*}_{j}^{1,L} } \cdot \alpha^{{s^*}_{j}^{1,L}} \mod p$ also holds. Therefore, \F~computes ${x^*}_{j}^{\ell} \as \hsim(x_j^* \| \ell, 0, l, \lh), \forall \ell \in [1,L]$ and calculates ${e^*}_{j}^{1,L} \as \sum_{\ell=1}^{L} { \hsim( {m^*}_{j}^\ell \| {x^*}_{j}^\ell, 2, l, \lh) } $. Thus, the following equations hold: 
		
		\vspace*{-4mm}
		\begin{eqnarray*}
			R_{j}^{1,L} \equiv Y^{e_{j}^{1,L}}\cdot \alpha^{s_{j}^{1,L}} \bmod p, ~~
			R_{j}^{1,L} \equiv Y^{{e^*}_{j}^{1,L}}\cdot \alpha^{{s^*}_{j}^{1,L} } \bmod p,
		\end{eqnarray*}

		\F~then extracts $y'=b$ by solving the below modular linear equations (note that only unknowns
		are $y'$ and $r_{j}^{1,L}$), where $Y=B$ as defined in the public key simulation:
		\vspace*{-2mm}
		\begin{eqnarray*}
			r_{j}^{1,L} \equiv y'\cdot e_{j}^{1,L} + s_{j}^{1,L} \bmod q~,~r_{j}^{1,L} \equiv y' \cdot {e^*}_{j}^{1,L} + {s^*}_{j}^{1,L} \bmod q,
		\end{eqnarray*}
		
		$B'\equiv \alpha^{b} \bmod p$ holds as  \A's forgery is valid and non-trivial on $B'=B$. By Def. \ref{def.dlp}, $\mathcal{F}$ wins the $\mathit{DL\mhyphen experiment}$.
	\end{list}
\end{enumerate}

\vspace{2mm}
\noindent \underline{{\em Execution Time Analysis}}: The runtime of \F~is that of \A~plus the time to respond to \ro~queries. We denote the approximate cost of drawing a random number and modular exponentiation with $\mathcal{O}(\kappa)$ and  $\mathcal{O}(\kappa^3)$, respectively.  In the setup phase, \F~draws random numbers with a negligible cost. In the query phase,  \F~performs $\mathcal{O}(3 \cdot L \cdot \kappa)$ to handle $\asig_{\sk}(.)$~oracle queries. To answer \comconstrmsk~ queries, it performs $ \mathcal{O}(2 \cdot L \cdot \kappa)$ (seed and ephemeral key derivation), and $\mathcal{O} (2 \cdot \kappa^3)$ for two modular exponentiations. The overall cost of query phase is bounded as $\mathcal{O}(\kappa^3)$.  Therefore, the approximate total running time of \F~is $t'=\mathcal{O}(t)+\mathcal{O}(\kappa^3)$.

\vspace{2mm}
\noindent \underline{{\em Success Probability Analysis}}: 
\F~succeeds if all below events occur.

\begin{enumerate}[-]
	\setlength{\itemsep}{0pt}
	\setlength{\parskip}{0pt}
	\setlength{\parsep}{0pt}
	\item \nab: \F~does not abort during the query phase.
	
	\item \forge: \A~wins the \AEUCMA~experiment for \lahases.
	
	\item \nabb: \F~does not abort after \A's forgery.
	
	\item   \suc: \F~wins the \AEUCMA~experiment for \dl{\em-experiment}.
	
	\item  $Pr[\suc] = Pr[\nab]\cdot Pr[\forge|\nab]\cdot Pr[\nabb|\nab \wedge \forge]$
	
\end{enumerate}

$\bullet$ {\em The probability that event \nab~occurs}: During the query phase, \F~aborts if $(m_j^\ell \| x_j^\ell, k=0)$ $\in\lh,\forall \ell \in [1,L]$ holds, {\em before} \F~inserts $(m_j^\ell \concat x_j^\ell, k=0)$ into \lh. This occurs if \A~guesses $x_j^\ell$ (before it is released) and then queries $(m_j^\ell \concat x_j^\ell)$ to \ro~{\em before} querying it to $\lahasessig(.)$. The probability that this occurs is $\frac{1}{2^{\kappa}}$, which is negligible in terms of $\kappa$. Hence, $Pr[\nab]=(1-\frac{1}{2^{\kappa}})\approx 1$.

$\bullet$ {\em The probability that event \forge~occurs}: If \F~does not abort, \A~also does not abort since the \A's simulated  view is {\em indistinguishable} from \A's real view. Thus, $Pr[\forge|\nab]=\advlahases$.

$\bullet$ {\em The probability that event \nabb~occurs}: \F~does not abort if the following conditions are satisfied:
(i) \A~wins the \AEUCMA~experiment for \lahases~on a message $M^{*}$ by querying it to \ro. The probability that \A~wins without querying $M^{*}$ to \ro~is as difficult as a random guess.
(ii) After \F~extracts $y'=b$ by solving modular linear equations, the probability that $Y' \not\equiv \alpha^{y'} \bmod p$ is negligible in terms $\kappa$, since $(Y=B) \in \pk$ and $\lahasesver(\pk,M^{*},\sigma^{*})=1$. Hence, $Pr[\nabb|\nab \wedge \forge]=\advlahases$.
Omitting the terms that are negligible in terms of $\kappa$, the upper bound on {\em \AEUCMA-advantage of \lahases}~is: ~~ $\advlahases \le \advdll$.

\vspace{2mm}
\noindent {\em \underline{Indistinguishability Argument}}: The real-view of \Areal~is comprised of the public key \pk, system-wide parameters $I$, the answers of $\lahasessig_{sk}(.)$ and $\lahases.\comconstrmsk$~(recorded in \ls~and \lc~by \F, respectively), and the answer of \ro~(recorded in \lh~and \lr~by \F).  All these values are generated by \lahases~algorithms as in the real system, where $sk=y$ serves as the initial randomness. The joint probability distribution of \Areal~is random uniform as that of \sk. 

The simulated view of \A~is as \Asim, and it is equivalent to \Areal~except that in the simulation, the signature components $\{s_j^\ell\}_{j\in[1,J], \ell\in[1,L]}$ are randomly selected from $\mathbb{Z}_q^*$. 
The private key is simulated as $y=z$, from which we derive the seeds $\{x_j\}_{j\in[1,J]}$ and the ephemeral keys $\{e_{j}^{1,L} \}_{j\in[1,J]}$. 
This dictates the selection of the aggregate signatures $\{s_j^{1,L}\}_{j\in[1,J]}$  and the aggregate commitments $\{R_j^{1,L}\}_{j\in[1,J]}$ as random via $\asig_{sk}(.)$~and \comconstrmsk~oracles, repectively.
Note that the joint probability distribution of these values is also randomly and uniformly distributed and is identical to the original signatures and hash outputs in \Areal~(since $H_{k=0,1}$ are modeled as \ro~via \hsim). 
\hfill $\blacksquare$

\vspace{2mm}
\noindent \underline{{\em On the Forking Lemma for Special-Condition in Schnorr-like}} \underline{\em Signatures}: 
Schnorr \cite{SchnorrQ} employs the forking lemma in its proof with a tight bound proven \cite{SecProofSigScheme96Euro}. Nonetheless, one-time Schnorr-like signatures (e.g., \cite{Yavuz:2012:TISSEC:FIBAF}) offer an alternative approach that eliminates the need for the forking lemma, but with the cost of pre-determined number of signing and linear public key size (or verifier transmission \cite{Yavuz:CNS:2019}). 
This trade-off arises because the commitment (i.e., $R$) is pre-fixed in \pk, meaning that the adversary \A~must forge using same commitment without the need to rewind the tape \cite{SecProofSigScheme96Euro}. 
Inline with these previous works, \lahases~generates randomness during signing, derived from initial randomness established during key generation. \lahases~introduces the concept of \cco~to provide verifiers with commitments in offline or on-demand. When an adversary \A~attempts forgery, they are supplied with aggregate commitments from $\comconstr_{\msk}(.)$. Consequently, \lahases~eliminates the need for the forking lemma, resulting in a more robust security guarantee.


\vspace{4mm}
We provide the security proof of \hyhases~as below.

\begin{theoremp}{5.3} \label{theorem3}
	$ \advhyhases=\min \{ Adv_{\pqhases}^{\EUCMA}(t, q_s, q_s',1)~,~ Adv_{\pqhases}^{\EUCMA}(t, q_s, q_s') \} $.
\end{theoremp}

\noindent {\em Proof:}
Let \A~finds a forgery against  $\hyhases$ with a nested combination of \pqhases~and \lahases~by  outputting $q_s +1$ valid \hyhases~signatures under distinct message batches. We can then construct an algorithm $\mathcal{A}_{LA}$ that finds a forgery in \lahases. $\mathcal{A}_{LA}$ interacts with  $\mathcal{F}_{LA}$ for \lahases~which provides a public key $\pk_{LA}$. $\mathcal{A}_{LA}$ derives a user's key pair $(\sk_{PQ},\pk_{PQ})$ by first generating $(\msk_{PQ}, \vec{sk}_{PQ},I)\as \pqhaseskg(1^\kappa, J, L)$. It sets the user's public key of \hyhases~to be $\pk_{HY} \as \langle \pk_{LA}, \pk_{PQ} \rangle$. 
When \A~asks for a batch of messages $\vec{M}_j = \{m_{j}^{\ell}\}_{\ell=1}^L$ to be signed using \hyhases, $\mathcal{A}_{LA}$ computes the nested vector $\vec{N}_j = \{n_{j}^{\ell}\}_{\ell=1}^L$, where $n_{j}^{1}=H_0(m_{j}^{1})$ and  $n_{j}^{\ell}=H_0(m_{j}^{\ell} \| H_0(n_{j}^{\ell-1}) )~,~\forall \ell=2,\ldots, L$. $\mathcal{A}_{LA}$ computes the signatures $\sigma_{LA,j}^{1,L}$ and $\sigma_{PQ,j}$ by passing $(\vec{N}_j,t)$ and $(n_j^L \| s_{LA,j}^L, t)$ to $\lahasessig_{\sk}$ and \pqhasessig, respectively.

If \A~wins the \HEUCMA~experiment, then it returns $q_s+1$ valid signatures $\sigma_{HY,j}^*=({\sigma^*}_{LA,j}^{1,L}, \sigma_{PQ,j}^{*})$ on distinct batches $\vec{M_j}$ such that $\lahasesver(\pk_{LA}, \vec{N_j}, {\sigma^*}_{LA,j}^{1,L})=1$ and  $\pqhases(\pk_{PQ,j}, n_{j,L} \| {s^*}_{LA,j}^L)=1$. As a result, $\mathcal{A}_{LA}$ can extract $q_S+1$ valid signatures under \lahases~on distinct messages. Thus, $\advhyhases \le \advlahases$ . The same result applies to \pqhases: $\advhyhases \le \advpqhases$. We conclude that:

$\advhyhases \le \min \{ \advlahases~\\, ~\advpqhases \}$
\vspace{3pt} 

$\advhyhases \le \min \{ J \cdot \advhyhors~,~\advhydll \}$
, where $t_{LA}'=\mathcal{O}(t)+\mathcal{O}(\kappa^3)$ and $t_{PQ}'=\mathcal{O}(t)+k \cdot H_0$.
\vspace{2mm}
\hfill $\blacksquare$

\end{document}